# Solid Oxide Electrolysis Cells: Bridging Materials Development and Process System Engineering for Gigawatt-Scale Applications

Matthias Riegraf,[a],* Marc Riedel,[b] Søren Hojgaard Jensen,[c,d] Srikanth Santhanam,[e] S. Asif Ansar,[a] Marc Heddrich,[a]

[a] German Aerospace Centre (DLR), Institute of Engineering Thermodynamics, Pfaffenwaldring 38-40, 70569 Stuttgart, Germany

[b] Robert Bosch GmbH, Robert-Bosch-Campus 1, 71272, Renningen, Germany

[c] Dynelectro ApS, Syvvejen 10, Hall 3, 4130 Viby Sjælland, Denmark

[d] Aalborg University, Department of Energy, Pontoppidanstræde 111, 9220 Aalborg, Denmark

[e] Shell Global Solutions International B.V, Grasweg 39, 1031 HW Amsterdam, Netherlands

*Corresponding author: E-mail: Matthias.Riegraf@dlr.de

## Abstract

High-temperature solid oxide electrolysis cells (SOECs) are a potential core power-to-X (P2X) technology due to their unparalleled system efficiencies, that can exceed 85 % when excess heat from exothermic downstream processes is available. Recent advancements in materials, cell and stack design have enabled the deployment of megawatt (MW) scale demonstration plants and gigawatt (GW) scale manufacturing capacities. Consequently, key challenges to industrial-scale adoption scale now increasingly lie at the system level. Unlike previous SOEC reviews focused on materials and stack-level innovations, this work uniquely addresses emerging interdisciplinary system-level challenges and highlights the need for a paradigm shift. Several key insights are identified. Pressurized operation plays a crucial role in enhancing SOEC system performance and enabling better process integration. The dynamic capabilities of SOECs are better than often assumed and can further be improved via advanced operating strategies and modularization. Balance-of-plant (BoP) component costs rival stack capital expenditure, emphasizing the need for cost reductions through economies of scale via mass production and cross-industry synergies. Co-electrolysis remains at a lower technology readiness level and lacks MW scale demonstration. Furthermore, demonstrated integration with downstream processes across entire P2X chains remains scarce. Future research and development strategies are proposed, offering a roadmap to overcome these challenges and accelerate SOEC commercialization.

## Broader context

Achieving climate neutrality and energy independence requires a rapid shift from fossil fuels to sustainable energy carriers. While renewable electricity is central to this transition, full decarbonization depends on sector coupling across power, industry, transport, and heating. Power-to-X (P2X) technologies enable this by converting renewable electricity into storable fuels and chemicals. Electrolysis is the core technology of P2X solutions, and solid oxide electrolysis cells (SOECs) are particularly promising due to their high efficiency and flexibility in producing hydrogen and syngas from $CO_2$ and $H_2O$. Their integration with exothermic industrial processes such as ammonia and methanol synthesis, which have been identified as high-priority use cases for hydrogen, offers unique synergies by using waste heat for steam generation. However, widespread industrial adoption depends on overcoming system-level challenges such as reducing system capital expenditure, ensuring operational flexibility, and achieving efficient integration with downstream processes. The present work identifies specific emerging interdisciplinary research challenges and promotes decarbonization of hard-to-abate sectors by accelerating SOEC market deployment.

## 1 Introduction

Hydrogen, due to its versatility as an energy carrier, is positioned to play a pivotal role in the global transition towards a low-carbon energy system. There remains considerable uncertainty about where hydrogen demand will materialize most strongly and green hydrogen is expected to remain scarce until at least 2040,[1] necessitating its prioritization for applications with the highest emissions abatement and economic benefit. High-priority use cases include heavy industry, fertilizer production, steelmaking, shipping and aviation.[2, 3] Many large industrial sites are currently not co-located with regions of abundant renewable electricity, and hydrogen transport remains costly.[2] Therefore, high-efficiency electrolysis technologies are a strategic necessity to ensure that limited clean electricity resources are used to maximize global decarbonization impact.

As electrolyzer capital expenditures (CAPEX) decrease through scale-up and learning, the levelized costs of hydrogen (LCOH) for all electrolysis technologies are projected to be increasingly dominated by the cost of electricity,[4, 5] making technologies that minimize electrical consumption particularly promising for industrial application. Solid oxide electrolysis cells (SOEC) offer electrical efficiencies up to 20 percentage points higher than those of low-temperature electrolysis technologies due their high operating temperatures of 550-850°C, which provide kinetic and thermodynamic advantages.[6] Efficiency gains are particularly pronounced when steam can be generated from available waste heat such as from strongly exothermic chemical synthesis processes. Moreover, SOEC allow efficient reversible operation and have the unique capability to operate in steam electrolysis, dry $CO_2$ electrolysis and co-electrolysis of $CO_2$ and $H_2O$, making high-temperature electrolysis attractive for a number of power-to-X (P2X) concepts.

The increasing interest in high-efficiency electrolysis technologies such as SOECs is also driven by supportive policy frameworks aimed at accelerating hydrogen deployment, particularly in the European Union. Initiatives such as REPowerEU and the updated Renewable Energy Directives (RED II and III) promote the adoption of renewable hydrogen technologies, specifically hydrogen produced from electrolysis using renewable electricitys, which is classified as a renewable fuel of non-biological origin (RFNBO).

Significant progress has been made in the development of solid oxide electrolysis cells and stacks in recent years. This resulted in scale-up now occurring at a rapid rate, building of manufacturing capacities up to the GW level and the installation of large demonstration plants on the MW scale. These developments have exposed new challenges on the system level, necessitating innovative concepts and strategies to address them. While the field has historically been driven by innovations in material science, the transition to large-scale systems and the widespread ongoing commercial implementation now increasingly demands advances in process system design, engineering and integration, fostering a more interdisciplinary integration of various scientific and technical disciplines. Compared to low-temperature electrolysis technologies, SOEC systems show a stronger cost dependency on scale due to the higher cost share of peripheral system components with improved scalability.[7, 8] These balance-of-plant (BoP) costs can be significantly reduced by centralization and optimization, spreading expenses over greater hydrogen output. As a result, system design and integration strategies that effectively leverage scale effects are not only technically important but also economically critical for making SOEC a competitive option in the market.

For this reason, this review will provide a concise overview of thermodynamic fundamentals, state-of-the-art SOEC materials, stack technologies and their predominant degradation mechanisms from a process systems perspective. We then connect these findings to critical aspects of SOEC system development and system challenges that need to overcome for a successful market introduction. This includes system design concepts, the development of suitable BoP components, process system aspects of scaling/numbering up strategies, downstream/upstream integration, and system operation and control that effectively mitigate challenges inherent to SOEC, such as degradation and mechanical fragility. These challenges are anticipated to become increasingly important as SOEC installations approach the GW scale. Unlike previous reviews that focused primarily on materials development, electrochemical performance and degradation,[9-11] this work provides a unique system-level perspective aimed at addressing the increasingly interdisciplinary challenges emerging from the scale-up of SOEC technology. This review is conceived as an initial holistic conceptualization of the still emerging field to help defining a future research agenda for the continued advancement of the field. By integrating both fundamental principles and practical applications, this review seeks to bridge the gap between scientific discovery and engineering implementation, offering a valuable resource for researchers across diverse disciplines, industry professionals and end-users preparing for SOEC integration at the multi-MW to GW scale.

## 2 Thermodynamic fundamentals

In an SOEC, $H_2O$ and $CO_2$ can be electrochemically reduced at the fuel electrode, generating oxygen ions that migrate through the electrolyte and form molecular $O_2$ at the oxygen electrode (Figure 1). The total energy demand is given by

$$\Delta H = \Delta G + T\Delta S \quad (1)$$

where $\Delta G$ represents the Gibbs free energy change, and $T\Delta S$ is the entropic term. Both water and carbon dioxide reduction are endothermic ((Figure 1b+c). For water splitting above the boiling point of water, the enthalpy of reaction is reduced by the enthalpy of evaporation, with a weak temperature dependence of $\Delta H$ (Figure 1b). $\Delta H$ for $CO_2$ reduction remains nearly constant across the entire temperature range considered (Figure 1c). As temperature increases, $\Delta G$ of both reactions decreases, while the entropic contribution ($T\Delta$S) increases, allowing a progressively larger share of the enthalpy to be supplied as heat.

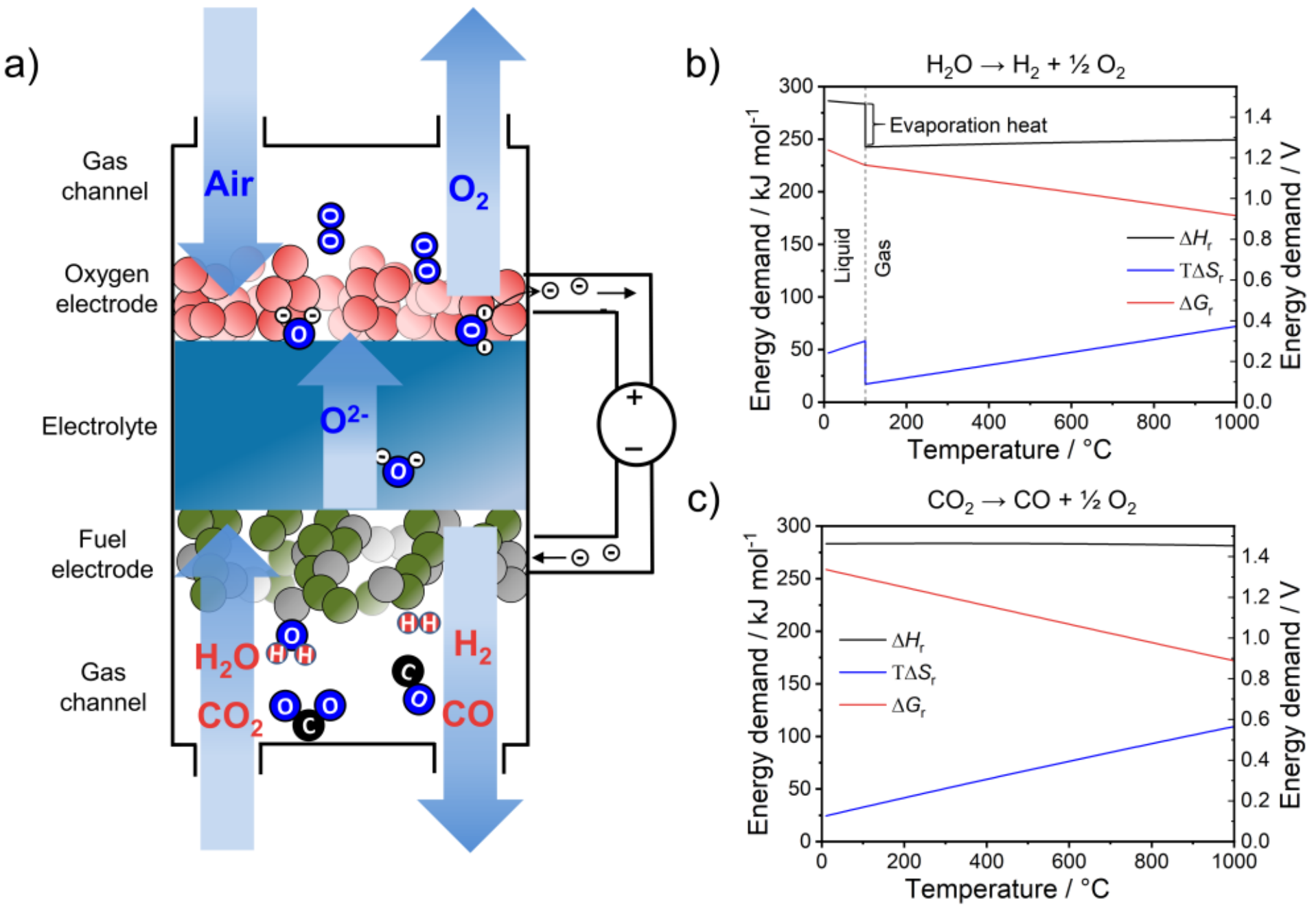

Figure 1. (a) Schematic illustration of the working principle of a solid oxide electrolysis cell. Thermodynamics of (b) $H_2O$ and (c) $CO_2$ electrolysis over temperature at 1 atm.

The reversible voltage $U_{\mathrm{rev}}$ is calculated according to

$$U_{\mathrm{rev}}=\frac{\Delta G}{\mathrm{nF}}, \qquad (2)$$

where $n$ is the number of transferred electrons ($n$ = 2 for $O^{2-}$) and F is the Faradaic constant, and decreases with temperature. Under ideal conditions, the reversible voltage is equal to the open-circuit voltage (OCV). The thermoneutral voltage

$$U_{\mathrm{tn}}=\frac{\Delta H}{nF}. \qquad (3)$$

is another characteristic cell voltage and corresponds to operation without heat exchange, where Joule heating balances the endothermic heat demand of the electrolysis reaction. At 800°C, $U_{\mathrm{tn}}$ is 1.29 V for steam electrolysis and 1.46 V for $CO_2$ electrolysis. Operaton above $U_{\mathrm{tn}}$ lead to exothermic operation, while operation below $U_{\mathrm{tn}}$ is endothermic.

During co-electrolysis, additional chemical reactions occur, such as the reverse water-gas shift (RWGS) reaction

$$H_2 + CO_2 \rightarrow H_2O + CO \qquad \Delta H\,(800°C) = 34\ \mathrm{kJ{\cdot}mol^{-1}} \qquad (4)$$

and methanation

$$4H_2 + CO_2 \rightarrow 2H_2O + CH_4 \qquad \Delta H\,(800°C) = -191\ \mathrm{kJ{\cdot}mol^{-1}} \qquad (5)$$

$$3H_2 + CO \rightarrow H_2O + CH_4 \qquad \Delta H\,(800°C) = -225\ \mathrm{kJ{\cdot}mol^{-1}} \qquad (6)$$

The stack efficiency $\eta$ of an electrolyzer is defined as:

$$\eta = \frac{U_{tn}}{U}, \qquad (7)$$

where $U$ is the operating voltage. Stack efficiency $\eta$ reaches 100% at thermoneutral voltage $U_{tn}$ and can even exceed 100% during endothermic operation. However, operation at thermoneutral voltage is generally favored as it minimizes temperature gradients inside the stacks. The overpotential at thermoneutral voltage $\eta_{\text{tn}} = U_{\text{tn}} - U_{\text{eq}}$ increases with temperature and is equal to the gap between $\Delta H$ and $\Delta G$ (Figure 1).

# 3 Cells and Stacks

Solid oxide cells (SOC) and stacks in various configurations and numerous materials are being explored in research. This review does not aim to provide an extensive overview of the ongoing progress in material, cell and stack development, since dedicated review articles have covered these topics in detail.[9, 10] Rather, this chapter briefly summarizes the state-of-the-art solutions that are currently considered for commercial application since their properties and behavior can have crucial implications for system design and operation.

## *3.1 Cell geometries and architectures*

SOEC development has largely built on the historical advancements in solid oxide fuel cell (SOFC) design, resulting in shared materials and architectures, and enabling reversible cell operation. The planar design predominates, offering advantages in cell interconnection and high current densities, although maintaining adequate sealing for prolonged operation is challenging. In contrast, tubular designs provide high mechanical durability and effective sealing also at high pressures, but face challenges with current collection and higher area-specific resistances.[12] Since SOECs are typically operated at lower thermal gradients than SOFCs, mechanical robustness against thermal stress is less critical, making tubular designs less common.

Irrespective of geometry, cell architecture hinges on a structural layer ensuring mechanical integrity, typically categorized as electrolyte-supported cells (ESC), fuel electrode-supported cells (FESC), and metal-supported cells (MSC) (Figure 2). ESCs employ thick electrolytes (60–150 µm), requiring high operating temperatures (750–850 °C) to achieve adequate ionic conductivity. Conversely, operating temperatures in FESCs (600-750°C) and MSCs (550-650°C), can be substantially reduced due to electrolyte thicknesses of 5–15 µm. FESCs face challenges from thick porous cathode supports, limiting tolerance towards thermal and redox cycling. MSCs utilize cost-effective and mechanically robust ferritic stainless-steel supports, offering enhanced thermal conductivity and rapid thermal cycling capability. However, the refractory properties of ceramic layers pose difficulties in densifying zirconia-based electrolytes during high-temperature sintering without undesired alteration of the metal substrate. Consequently, electrolyte and electrodes must be fabricated at restricted temperatures, potentially impacting their performance.

Overall, differences between architectures reflect application-specific optimization rather than the absence of an ideal design.

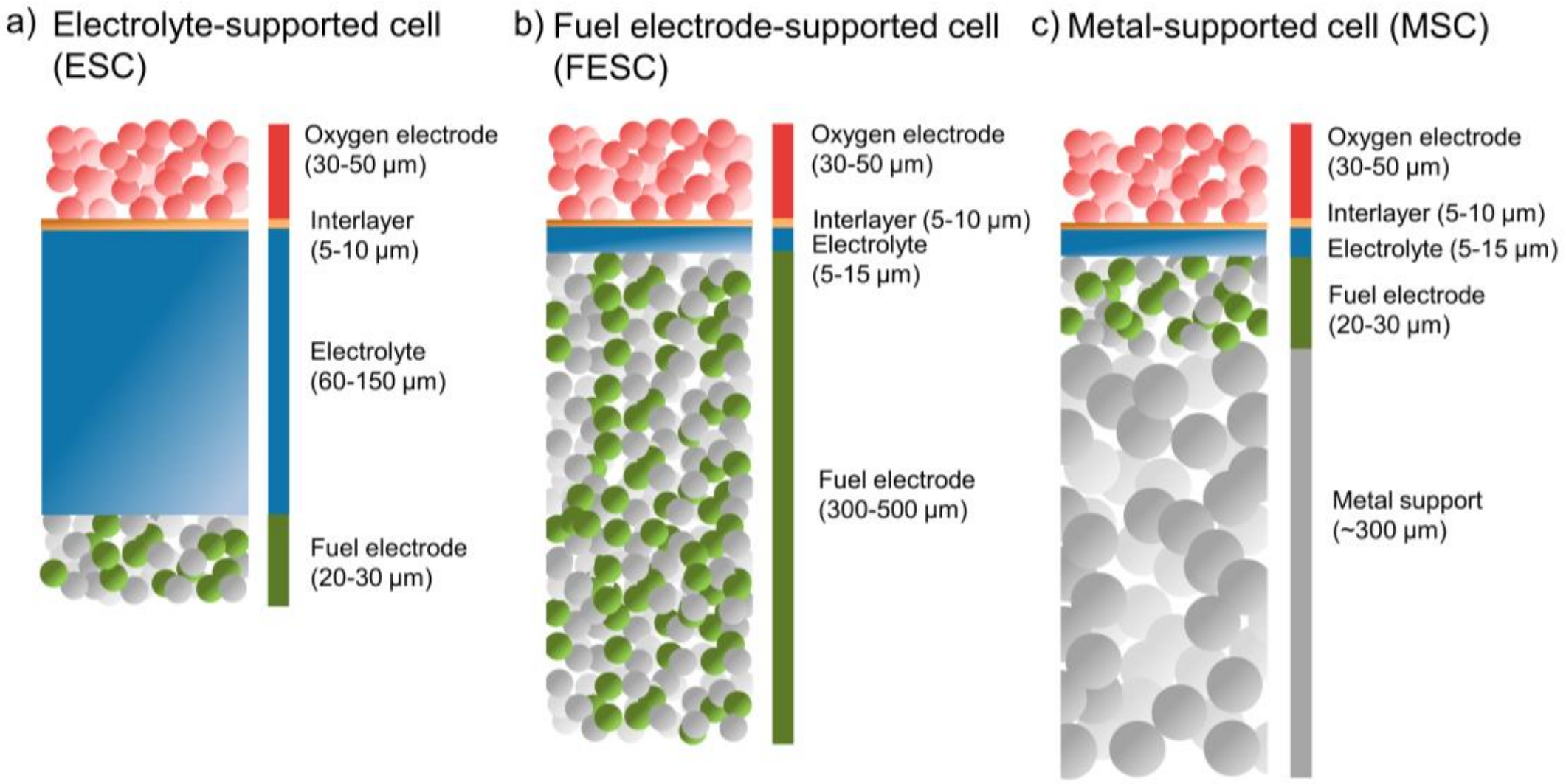

Figure 2. Schematic illustration of the different planar SOEC designs: (a) electrolyte-supported cell, (b) cathode (fuel electrode)-supported cell, and (c) metal-supported cell.

### *3.2 Electrolyte*

State-of-the-art electrolyte material is zirconia ($ZrO_2$) doped with yttria ($Y_2O_3$) and scandia ($Sc_2O_3$) which leads to extrinsic ionic transport properties.[13] Compositions with higher dopant concentrations offer increased conductivity at the expense of mechanical strength. Yttria-doped zirconia (YSZ) is cost-effective and abundant, while scandia-doped zirconia (ScSZ) offers superior performance but scandia is less available and more expensive.[14, 15] Alternatively, Gadolinium-doped ceria (CGO) can be used as electrolyte material which shows particularly high ionic conductivities at temperatures below 700°C. However, its mixed ionic and electronic conductivity (MIEC) properties and chemical expansion in reducing atmosphere at higher temperatures restricts its use to temperatures of 600°C and lower.[16]

### *3.3 Oxygen electrode*

Modern SOEC oxygen electrodes mainly consist of perovskites with mixed ionic electronic conductivity (MIEC) such as $La_{0.6}Sr_{0.4}Co_{0.2}Fe_{0.8}O_{3-\delta}$ (LSCF) and $La_{0.6}Sr_{0.4}CoO_{3-\delta}$ (LSC).[17]

They display high electronic conductivities, electro-catalytic performance and favorable oxygen transport properties. They can also be combined with CGO to further increase their apparent oxygen ion conductivity.[18, 19] To prevent the formation of insulating strontium zirconate interfacial layer at the oxygen electrode/doped zirconia electrolyte interface,[20] a doped ceria diffusion barrier layer (e.g. CGO) is commonly applied between YSZ and LSCF/LSC.[21-24]

### *3.4 Fuel electrode*

State-of-the-art fuel electrodes are porous composite electrodes containing a percolating nickel and a ceramic ionically conducting phase such as CGO or YSZ.[25] The Ni phase acts as electronic conductor and catalyst in the water splitting reaction, and has the additional benefit of catalyzing the RWGS reaction during co-electrolysis. In Ni/YSZ, the reaction is confined to the triple phase boundary (TPB),[26] whereas the reaction zone extends to the CGO surface due to its MIEC properties and electro-catalytic activity.[27] For this reason, Ni/CGO fuel electrodes show lower overpotentials and are preferred in ESC and MSC.[25] However, CGO cannot supply the high mechanical support strength required in FESC. For this reason, state-of-the-art FESC are generally based on Ni/YSZ supports.

### *3.5 Stack components and architectures*

#### *3.5.1 Interconnects*

Interconnects electrically connect cells in a stack while ensuring gas separation and distribution. They operate in harsh conditions, exposed simultaneously to oxidizing and reducing environments. As a result, they require high chemical stability, along with high electrical and thermal conductivity.

Ferritic stainless steels are widely used, with specialized grades (such as Crofer22APU) offering high conductivity, corrosion resistance and durability. At higher temperatures ~800°C where mechanical stress and creep become critical, oxide dispersion strengthened (ODS) alloys are another common choice due to enhanced corrosion resistance, strength and creeping resistance.[28, 29] Protective coatings such as conductive spinel oxides (e.g. $(Co,Mn)_3O_4$ (MCO)) are often applied to suppress chromium evaporation and further improve corrosion resistance. At lower temperatures (~600°C), as in MSC architectures, more cost-effective ferritic steels can be used, often without protective coatings due to reduced corrosion risk.[30]

*3.5.2 Sealants*

To avoid hydrogen dilution with oxygen, effective sealants are required in planar SOEC stacks. They either prevent leakage of gases from the stack to the environment, or the mixing of gases within the stack. Depending on the stack design, a number of different sealing concepts have been devised.[31] In all cases, they need to be gastight, mechanically stable, electrically insulating, and able to withstand high temperatures, as well as simultaneous oxidizing and reducing atmospheres. Furthermore, the sealing material needs to have a coefficient of thermal expansion (CTE) that matches that of both the electrolyte and the interconnect. For this purpose, glass-ceramic sealants such as alkali or alkali-earth silicate-based materials are most commonly used.[32] In MSC at reduced operating temperatures, simple compressive gaskets can provide a cost-effective alternative.[33]

*3.6 Stack designs*

SOEC stacks exist in several configurations, with the main features of commercial stack suppliers summarized in Table 1. Among them, Mitsubishi Hitachi Power Systems is the only company using a tubular design, although no details are publicly available. For this reason, the following discussion focuses on the planar design.

Early SOEC designs were derived from reversible SOFC stacks with only minor modifications. However, recent developments showed that stack requirements in both operating modes can be substantially different. For instance, while SOFCs demand high air flow rates for cooling, SOECs operate with much lower flows, and in extreme cases only with oxygen generated during electrolysis. This enables narrower oxygen electrode channels and thinner interconnects, reducing cost. Consequently, companies such as Sunfire and Topsoe have shifted towards stack architectures specifically optimized for SOEC operation.

Fundamental design differences remain, particularly in flow-field configurations (cross-, co-, counter or radial flow) and air supply (external supply in open air design or internal manifolding).

Table 1. Overview of different stack manufacturers, their employed cell architecture and composition, cell number, interconnect material, flow pattern, air manifold and active area. Since progress occurs at a fast rate, details are subject to changes.

| **Stack manufacturer** | **Cell architecture** | **Cell composition** | **Repeat units** | **Interconnect** | **Flow pattern** | **Air manifold** | **Active area / cm$^2$** | **Nominal operating temperature / °C** |
|---|---|---|---|---|---|---|---|---|
| Bloom Energy Corporation (USA) | ESC | Stabilized zirconia support [34] | N/A | 96 % Cr, 4 % Fe (assumed[35, 36]) | Cross-flow (assumed[37]) | N/A | N/A | N/A |
| Ceres Power Holdings plc (UK)[38, 39] | MSC | Ferritic steel\|Ni/CGO\|CGO\|YSZ\|CGO\|Perovskite | 300 | Ferritic steel | Co-flow [30] | N/A | 234 | ~550 |
| E&KOA Co. (Korea)[40-42] | FESC | Ni/YSZ\|8YSZ\|LSC from Elcogen | 120 | Ferritic steel type 460 FC from POSCO | Cross-flow | Internal | 100 | ~750 |
| Elcogen AS (Estonia)[43] | FESC | Ni/YSZ\|8YSZ\|LSC | 119 | Ferritic steel, coated with MCO | Co-flow | External | 121 | ~650 |
| Fraunhofer IKTS/Thyssenkrupp Nucera AG (Germany)[25, 28, 44-46] | ESC | Ni/CGO\|10Sc1CeSZ\|LSMM' | 40 | CFY | Cross-flow | External | 127 | ~800 |
| FuelCell Energy, Inc (USA)[47] | FESC | Ni/YSZ\|YSZ\|Perovskite | 350 | Ferritic steels | Radial | N/A | 81 | 700-750 |
| MiCoPower (Korea) | FESC (assumed[48]) | N/A | N/A | N/A | N/A | N/A | N/A | 700-750[42] |
| Mitsubishi Hitachi Power Systems, Ltd. (MHPS) (Japan) [49] | Tubular, gas-permeable ceramic substrate tube | N/A | N/A | Electron-conductive ceramic | N/A | N/A | N/A | N/A |
| Nexceris, Inc. (USA)[50] | ESC, honeycomb structure | Ni/Cermet\|YSZ\|LSCF | 36 | N/A | Cross-flow | N/A | 96-228 | N/A |
| Oxeon Energy, LLC (USA)[51] | ESC | Ni/cermet\|ScSZ\|Perovskite | 65 | CFY | N/A | External | ~169 | 750-850 |
| Posco Energy (Korea)[42] | FESC (assumed[42]) | N/A | 36 | Poss460FC | N/A | N/A | 100 | 700 |
| SOFCMAN (China) | FESC | Ni/YSZ\|YSZ\|LSCF/CGO | 50 | SUS430 | Co-flow | Internal | 97 | 700-750 |
| SolydEra SpA (Italy)[52, 53] | FESC | Ni/YSZ\|YSZ\|LSCF/CGO | 4x80 (window design)[54] | Crofer 22APU[55] | Co-flow | N/A | N/A | ~720 |
| Sunfire SE (Germany)[56, 57] | ESC | Old cell design: Ni/CGO\|3YSZ\|LSCF/CGO; New cell design: Ni/CGO\|ScSZ\|LSCF/CGO | Old stack design: 30 cells; New stack design: Unknown | Crofer 22APU | Co-flow | External | 127.8 | 820 |
| Topsoe A/S (Denmark)[58] | FESC | Ni/YSZ\|YSZ\|Perovskite | 50-150 | Ferritic stainless steel (assumed[59]) | N/A | N/A | Hexagonal, estimated 300-400 | ~750 |
| Vermes SOC Technology[60]/Chaozhou Three-Circle (Group) Co., Ltd. (CCTC) (Germany/China) | FESC | Ni/YSZ\|8YSZ\|LSC(F) | 51 | Ferritic steel | N/A | Internal or External | N/A | 650-700 |

## 4 Degradation mechanisms

Reliable long-term operation is crucial for the widespread adoption of SOEC, and substantial development efforts are focused on mitigating cell and stack degradation. Comparison across studies is complicated by the lack of standardized metrics and different operating conditions.[61] Most stack tests are carried out galvanostatically, with degradation typically expressed as relative voltage increase (%/kh) or area-specific resistance (ASR, in $m\Omega\ cm^2/kh$) increase. However, these values can depend strongly on current density, temperature, and gas composition.

The European Clean Hydrogen Joint Undertaking targets degradation rates of 0.5 %/kh by 2030 which remain higher than the 0.1-0.2 %/kh for alkaline electrolysis (AEL) and proton exchange membrane electrolysis (PEMEL).[62] Nevertheless, 0.5 %/kh is already routinely achieved for major stack technologies in steam electrolysis,[63-65] and values as low as 0.1 %/kh have been demonstrated.[66] Co-electrolysis introduces additional degradation mechanisms, but comparable degradation rates have been reported when harmful operating conditions are avoided.[28]

Degradation arises from various different intrinsic (e.g., material properties, surface and interface evolution) and extrinsic (e.g., external stressors such as contaminants) factors (Figure 3). These mechanisms must be considered during system design and can be partially mitigated through effective operating strategies. The most critical degradation mechanisms are briefly reviewed in the following, and more detailed discussions are available elsewhere.[61]

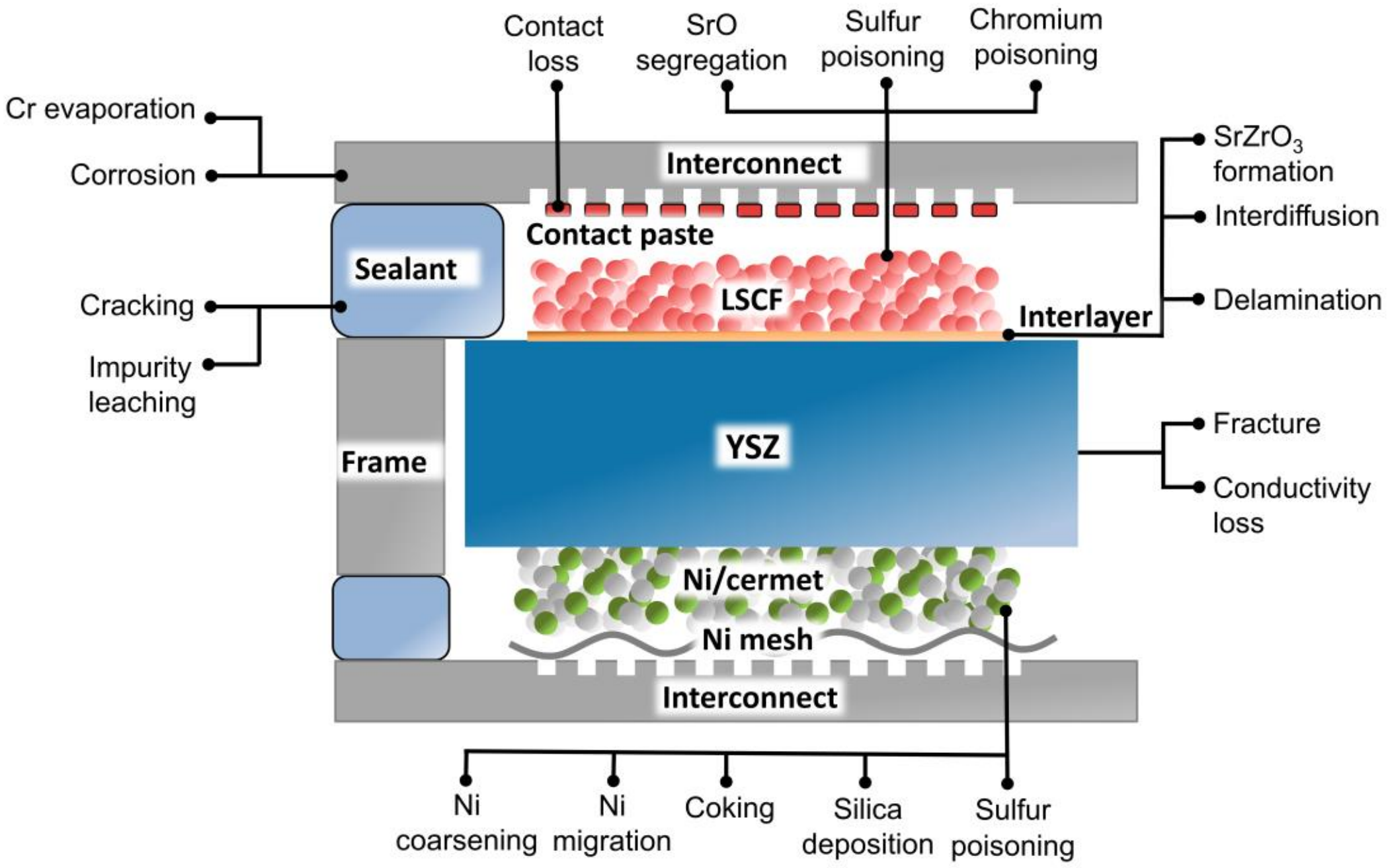

Figure 3. Overview of common degradation mechanisms in a schematic SOEC.

### 4.1 *Electrolyte*

In doped-zirconia electrolytes, gradual decreases in ionic conductivity lead to increased ohmic losses.[67-69] This is generally the dominating degradation process in ESCs due to their large electrolyte thickness,[63] whereas it plays a smaller role in the case of FESCs/MSCs with thinner electrolytes.

### 4.2 *Fuel electrode*

Ni is mobile at SOEC operating conditions with high temperature, humidity content and large cathodic bias which can lead to Ni coarsening reducing its surface area.[70, 71] Lowering the operating temperature can mitigate this issue.[72] Re-oxidation of Ni in the absence of reducing gas leads to dimensional changes and microstructural damage, particularly in Ni/YSZ,[73, 74] whereas Ni/CGO fuel electrodes tend to exhibit higher redox stability.[75]

Ni migration away from the electrode/electrolyte interface can be another prominent degradation process in Ni/YSZ fuel electrode via an effective lengthening of the ionic diffusion paths.[70, 76, 77] The onset of this degradation process is a complex function of temperature, humidity and fuel electrode overpotential. Ni migration can be the primary source of degradation in FESCs with Ni/YSZ fuel electrodes,[76] but has not been observed for Ni/CGO electrodes.

Ni is also susceptible to surface poisoning by fuel gas impurities. Sulfur impurities in the feed gas can cause significant performance drops during $CO_2$ electrolysis even in the ppb range, whereas co-feeding hydrogen during co-electrolysis strongly mitigates this effect.[78, 79] Silica deposition at the electrode/electrolyte interface can also lead to performance losses.[80] Possible silica sources are feed water, steel pipes, evaporators, electrical heaters, glass sealants, and the raw materials of the cell components.[80-83]

The formation of CO-rich gas mixtures during co-electrolysis and $CO_2$ electrolysis can lead to solid carbon deposition according to the Boudouard reaction

$$CO_2(g) + C(s) \rightarrow 2CO(g). \qquad (8)$$

Carbon formation is thermodynamically favored at low temperatures and high pressures, and the thick fuel electrode support layers in FESCs/MSCs can exacerbate carbon deposition due to local CO enrichment near the electrode/electrolyte interface.[78] Furthermore, electrochemical carbon formation was suggested to occur at high fuel electrode overpotentials.[84]

### *4.3 Oxygen electrode*

A number of different degradation processes are commonly reported for LSC(F) electrodes.[10] Their microstructures has been reported to be stable,[85] but SrO surface segregation can lead to surface passivation,[85, 86] particularly at high temperature and humidity.[10] The migration of volatile Sr species to the electrolyte/doped ceria interface can lead to the formation of secondary phases with low ionic conductivity such as $SrZrO_3$.[20, 87]

Furthermore, electrode degradation can occur due to the formation of secondary phases upon the reaction of segregated Sr with impurities such as sulfur and chromium.[88] A common origin of volatile Cr is the metallic interconnect where gaseous Cr species can form particularly in the presence of moisture.[89] Trace sulfur compounds are present in the ambient air at the ppb level and can further contribute to electrode decomposition.[90, 91]

### *4.4 Mechanical degradation*

The different layers in SOECs generally have different CTEs and are bonded together by high-temperature sintering, which creates a stress-free state during fabrication, but induces residual thermal stresses during operation. This can lead to a substantial reduction in mechanical strength, that is, the critical stress value.[22, 92] Additional stresses can arise during operation from thermal gradients within the stack, chemically induced strain (e.g. Ni oxidation in the fuel electrode, or reduction-induced expansion of doped ceria which restricts its use to ~600°C[93]), and externally applied stack compression. Such combined stresses can cause fast fracture and stack failure.

An initial mechanical strength analysis for temperature cycling will predict only either immediate mechanical failure or infinite mechanical stability. However, in reality most cells will fail after a certain number of thermal cycles due to time-dependent mechanical degradation phenomena, such as change in material properties (e.g. redistribution of Ni upon redox cycling), slow subcritical crack growth and creep.[93]

Since mechanical failure is typically initiated by large defects that are stochastically distributed, mechanical lifetimes cannot be precisely predicted solely from load, materials and geometry.

### *4.5 Interconnect and sealant*

In the absence of an effective protective coating, progressing chromia scale growth on the metallic interconnect increases its electrical resistance over time, and Cr evaporation causes severe oxygen electrode degradation.[94, 95] Both processes are affected by the operating conditions, and thus, protective coatings and interconnect choices need to be optimized for each stack design.[96, 97] So far, in most cases no protective coatings are applied on the hydrogen side and in MSC supports.[98-100]

High steam partial pressures can lead to the leaching of different elements such as silica into the fuel gas stream causing degradation of the glass sealant and secondary degradation effects in the fuel electrode.[81] Moreover, fast and repetitive thermal cycling as for instance, during reversible operation, can damage the sealants and compromise the gas tightness of the sealing due to mechanical stresses.[63] In addition, such fast temperature changes can also induce the risk of contact loss between cell and interconnect.[101]

## 5 System design and balance of plants (BoP) components

High-temperature operation introduces unique challenges with regards to heat management, material compatibility and system integration. Efficient system design including the BoP components must address these factors to maximize efficiency, reliability and economic viability. Multiple stacks and their support structure are grouped into a stack module, which is then assembled in a hotbox (Figure 4) that also includes piping for gas distribution and thermal insulation. Hot BoP components such as high-temperature (HT) heat exchangers (HEX), electric heaters and HT recirculation components are located in the direct vicinity of the hotbox. Multiple hotboxes are combined with HT and low-temperature (LT) BoP components (e.g. power electronics, steam generators and blowers) into a repeatable, self-contained SOEC system or unit (Figure 4). Depending on the design, a hotbox and a system may include the same number of stacks, while some components such as steam generators or hot recovery units (consisting of LT HEX) may be implemented in a centralized configuration at the plant level or in a decentralized configuration at the system level.

Modularization then enables scale-up by numbering-up to compose large SOEC plants,[102] with limited additional economic scaling advantages beyond the system level. The proposed nomenclature of stack module, hotbox and system/unit is not universally accepted, but this terminology is used throughout the present work. An exemplary process flow diagram for a (co-)electrolysis system coupled with a downstream synthesis reactor is depicted in Figure 5.

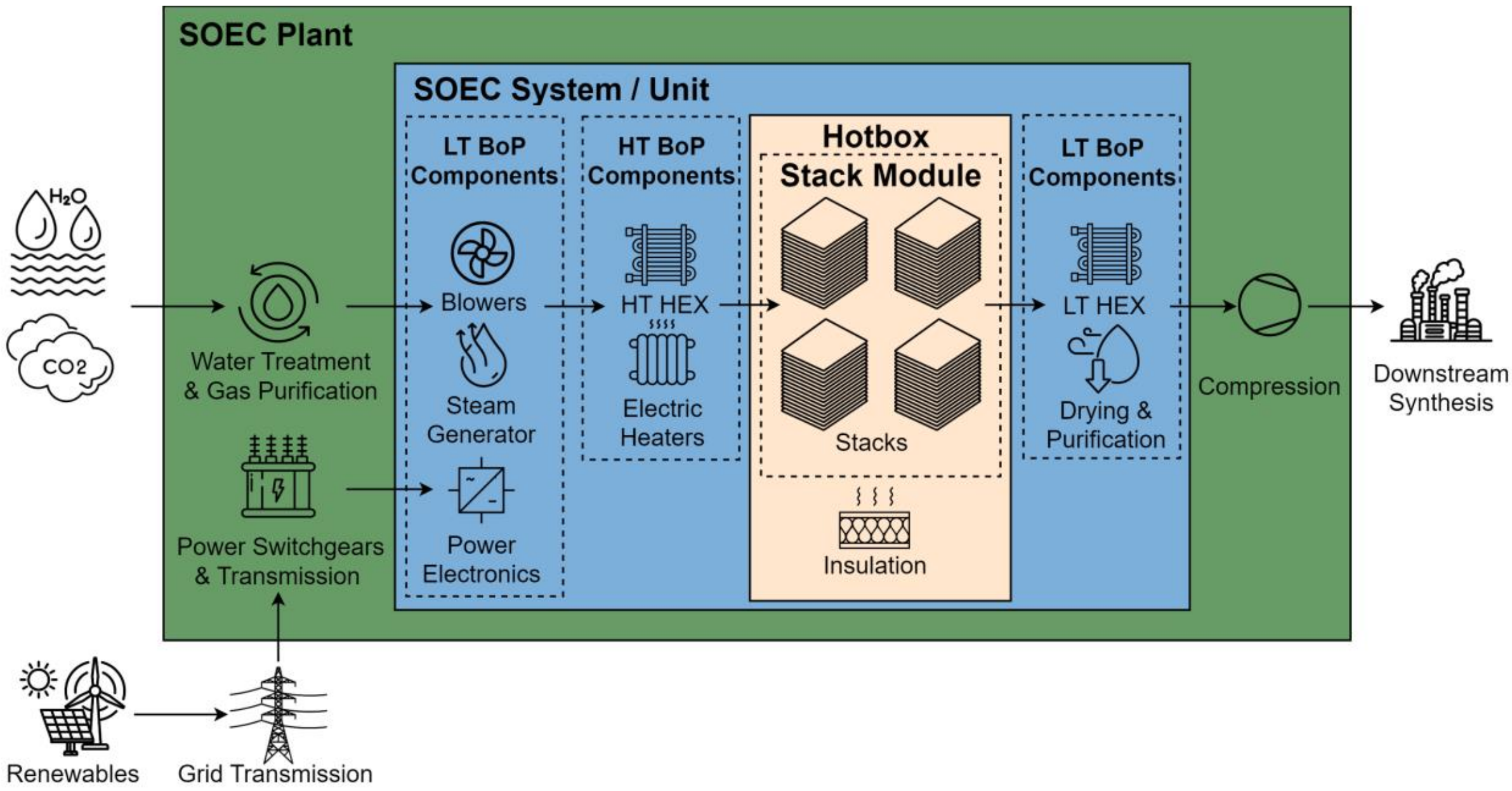

Figure 4. Schematic illustration of SOEC stack module, hotbox, system/unit and plant. Hotbox includes the multi-stack module, piping for gas distribution and thermal insulation. as well as. Multiple hotboxes are combined with the hot BoP components such as HT HEX and electric heaters, and the centralized LT BoP components (e.g. steam generators, power electronics and recirculation blowers) into a SOEC system/unit. Modularization enables scaling to compose large SOEC plants. Water treatment, gas purification and compression are generally considered to occur on plant level.

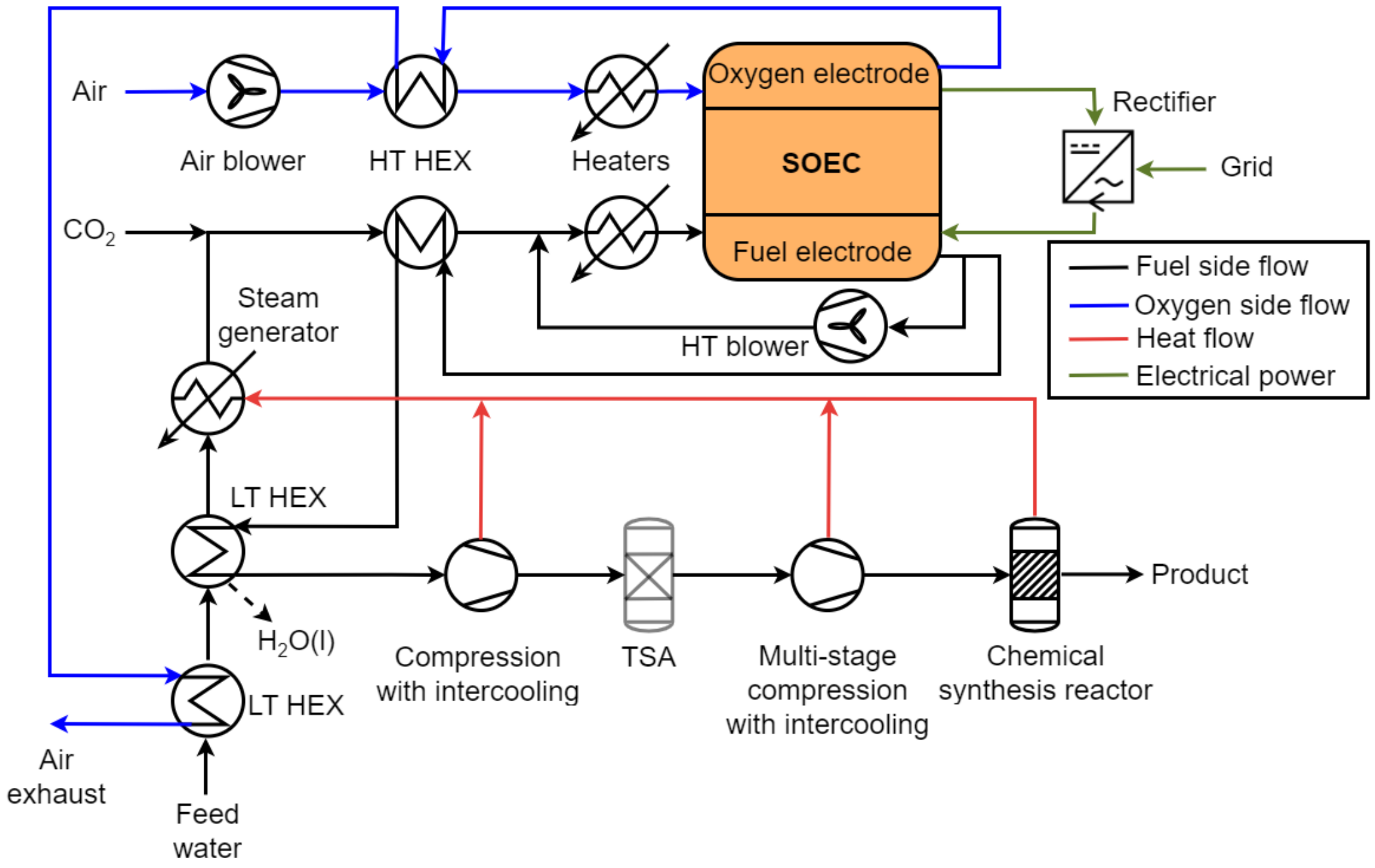

Figure 5. Simplified process flow diagram of a (co-)electrolysis system coupled with a chemical synthesis reactor. Fuel gas is recirculated with an HT blower, alternative recirculation strategies are explained in the following. The temperature swing adsorption (TSA) unit is depicted in low opacity to indicate its optionality depending on the coupled chemical synthesis reactor.

### 5.1 *System Efficiency*

While the stacks can be operated close to 100% stack efficiency (according to eq. 7), the overall system efficiency is reduced due to the energy demand of BoP components and thermal losses. The SOEC system efficiency is calculated according to

$$\eta_{LHV,sys} = \frac{LHV \cdot \dot{m}_{Prod}}{P}, \qquad (9)$$

with LHV being the lower heating value per mass, $\dot{m}_{Prod}$ the total mass flow of the products and $P$ the total electrical power input. Main loss contributions usually include hotbox heat losses to the environment, electric trim heaters, power electronics and, when required, electric steam evaporation.

Upon availability of steam, $\eta_{LHV,sys}$ exceeding 85% $_{LHV,AC}$ has been demonstrated, for instance by Sunfire already in 2019, corresponding to ~39 kWh/kg$_{H2}$.[56] More recent systems from Sunfire reportedly could achieve up to 88 %$_{LHV,AC}$, or 37.8 kWh/kg$_{H2}$.[103] Similarly, Bloom Energy and Ceres report ~37 kWh/kg$_{H2}$ specific electricity consumption in their systems.[104, 105] This is significantly lower than the reported specific energy consumptions of 52-55 kWh/kg$_{H2}$ and 49-50 kWh/kg$_{H2}$ for PEM electrolysis and alkaline electrolysis systems, respectively.[62]

Importantly, while end-of-life (EoL) efficiencies of low-temperature electrolysis systems typically decrease by 6-7 percentage points relative to beginning of life (BoL) values, no such decline is typically observed for SOEC, further enhancing its efficiency advantage (Table 2 and Figure 6).

If the heat of evaporation is entirely supplied by electricity, system efficiency values decrease by 12.5 percentage points or more (~6.6 kWh/kg$_{H2}$).[56] However, such cases are expected to be rare due to partial thermal steam generation from internal waste heat recovery. Assuming internal heat recovery covers one third of the steam demand,[106] state-of-the-art atmospheric SOEC systems could still achieve 77-80 % efficiency, significantly outperforming low-temperature electrolysis technologies even in the absence of an external heat source.

However, while low-temperature electrolysis systems are increasingly capable of operating at elevated pressures of ~30 bar,[103] state-of-the-art SOEC systems typically operate at atmospheric pressure. As a result, additional energy demand of ~2.8 kWh/kg$_{H2}$ is required to compress the produced hydrogen to 30 bar. This reduces system efficiency by ~5.8 percentage points.

Hence, the streamlined integration of the SOEC stack module with BoP components into a system that allows durable and efficient operation is key to the wide-spread adoption of solid oxide electrolysis. The following section will review the required BoP components and their role in optimizing system performance.

Table 2. Comparison of different SOEC and low-temperature electrolysis system efficiencies. Efficiency for thermal steam generation with excess heat is based on data from Ref.[56, 103] Losses for electric steam generation were taken from Ref.[56] The assumption of internal heat recovery covering 1/3 of steam demand was based on a recent study.[106] Specific energy consumption for compression to 30 bar were calculated assuming 4-stage compression with 80% compression efficiency. Low-temperature electrolysis values are given at BoL according to manufacturer data. EoL values were estimated on the economic determination of stack replacement after a ~15 % voltage increase,[107] which is assumed to correspond to a conservative 10% relative loss in system efficiency.

| System configuration | $\eta_{LHV,sys}$ / % | Specific energy consumption / kWh/kg$_{H2}$ |
|---|---|---|
| SOEC (full steam generation with excess heat, no compression) | 85-89 | 37-39 |
| SOEC (full electric steam generation, no compression) | 73-77 | 43-45 |
| SOEC (internal heat recovery leading to 1/3 thermal steam generation, no compression) | 77-81 | 41-43 |
| SOEC (steam generation with excess heat, compression to 30 bar*) | 79-83 | 40-42 |
| SOEC (full electric steam generation, compression to 30 bar) | 67-71 | 46-48 |
| SOEC (internal heat recovery leading to 1/3 thermal steam generation, compression to 30 bar) | 71-74 | 45-46 |
| AEL (30 bar) BoL/EoL | 61-64 / 55-58 | 52-55 / 58-61 |
| PEMEL (30 bar) BoL/EoL | 67-68 / 60-61 | 49-50 / 54-56 |

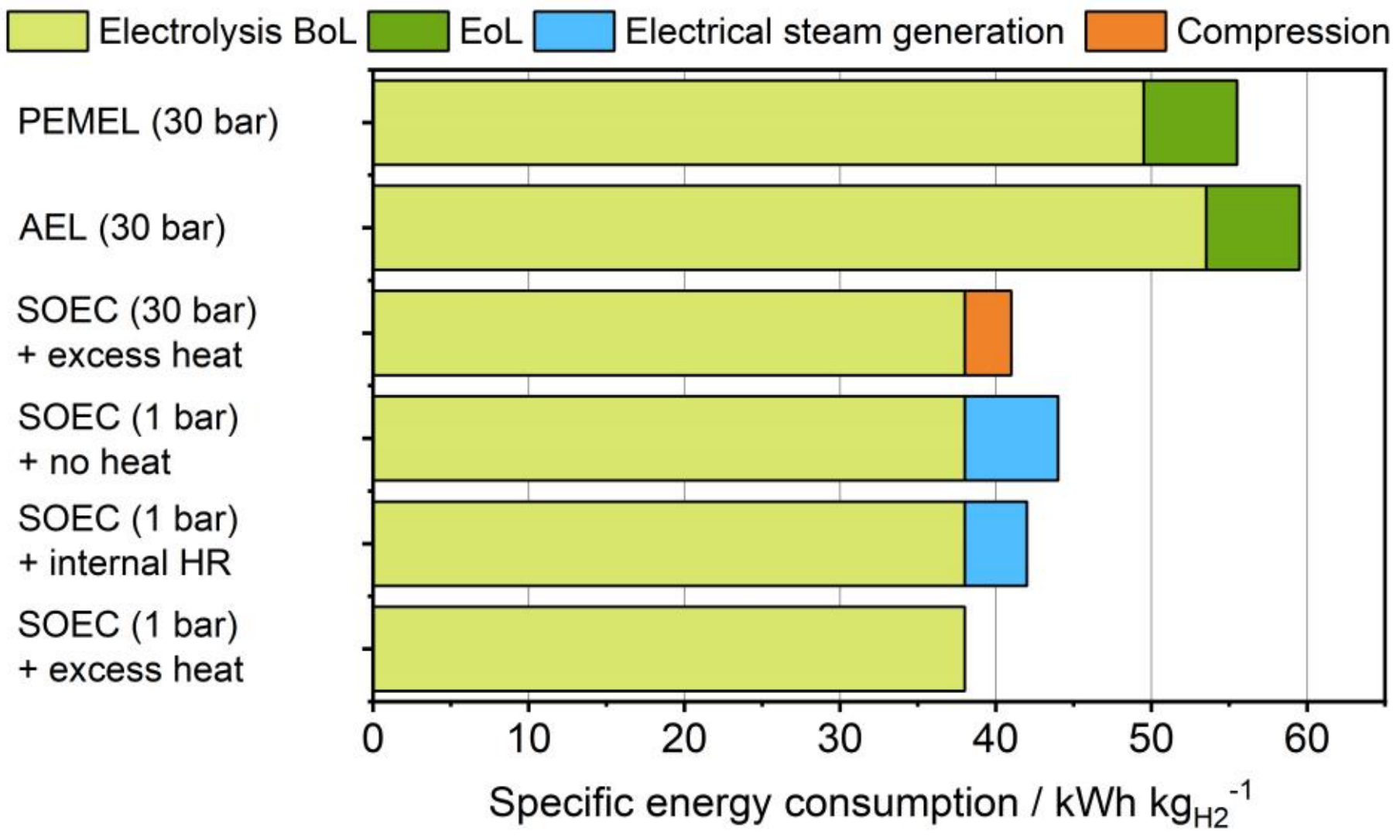

Figure 6. Comparison of specific energy consumption of different state-of-the-art SOEC and low-temperature electrolysis system efficiencies based on values given in Table 2. HR = (internal) heat recovery used for steam generation.

## *5.2 BoP components*

### *5.2.1 Water management and steam generation*

Besides electricity, water is the primary input for electrolysis, with fresh water being the state-of-the-art. Theoretically, 9 L of water is required to produce 1 kg of hydrogen;[108] corresponding to a water consumption of 230,000 L/h for a 1 GW SOEC system (with 38 kWh/kg$_{H2}$). Actual consumption is typically even higher. Fresh water resources are already under stress in many regions in the world, a challenge expected to intensify with political and economic consequences.[108] Limited global reactant conversion (RC) in SOEC leads to significant condensation of water downstream, and additional water is generated in coupled processes such as methanation, Fischer-Tropsch (FT) and methanol synthesis. This opens up the possibility of closed-loop systems where product water is recycled as SOEC feed, potentially reducing fresh water demand. However, impurity concentrations in the product water remain unknown and require further investigation.

Seawater desalination offers an alternative, particularly in coastal areas with abundant renewable energy. Reverse osmosis (RO) is used in ~70% of desalination plants and has minimal impact on hydrogen production costs.[109] However, brine discharge from desalination poses environmental risks due to its high salinity.

Even after RO treatment, fresh water with a typical conductivity <1000 μS $cm^{-1}$ is generally unsuitable for SOEC operation due to possible impurity deposition in the stack. Thus, advanced purification to ultrapure water (<0.1 μS $cm^{-1}$) is recommended.[80] In particular, silica is commonly present in many water sources due to mineral dissolution and should be removed, typically via mixed bed ion exchange resins.[80] Importantly, higher water purity requirements lead to increased volumes of reject streams during purification, which in turn raises overall fresh water demand.

Steam is often generated using waste heat in highly integrated systems, such as steam boilers or heat recovery steam generators (HRSG).[110] Ideally, HRSGs should accommodate multiple heat sources, as well as electricity. Steam can partially be generated by recuperating the heat from the off-gases via LT HEXs. Additional heat sources can come from the downstream reactors, utilizing the exothermic reaction heat and energy from compressors with intercooling. Steam generation and heat recovery units may be implemented centrally using downstream reactors, decentrally using local heat from SOEC off-gases, or as a combination of both, with the approach chosen based on system boundary conditions to optimize efficiency, modularity, and operational flexibility.

### *5.2.2 Feed gas supply*

For $CO_2$ or co-electrolysis operation, carbon dioxide could be extracted from fossil point sources such as cement plants by using carbon capture technology and then re-utilized to reduce total $CO_2$ emissions.[111] While $CO_2$ from fossil origin ultimately returns to the atmosphere, making it less suitable for long-term carbon neutrality, this pathway may serve as an interim strategy.[112] Fully carbon-neutral $CO_2$ could be captured from biogenic sources, such as biogas or biomass combustion, provided a sustainable biomass source. Carbon dioxide streams can also be generated via direct air capture (DAC) with an even lower carbon footprint,[113] but DAC faces challenges with regards to energy, material and land use.[112] Different carbon capture technologies can achieve high $CO_2$ purities, but higher purity requirements increase costs.[114] The impact of potential impurities on SOEC fuel electrodes remains insufficiently explored. For instance, many $CO_2$ sources contain sulfur, necessitating desulfurization in $CO_2$ electrolysis applications, whereas such cleaning may not be required for co-electrolysis. Since sulfur removal at ppb levels is difficult at ambient conditions,[79] high-temperature fuel gas cleaning before the inlet has been proposed.[115]

Purging the oxygen side with high flow rates of air is a widely used strategy for thermal management.[116-118] This approach also dilutes the produced oxygen which improves process safety and minimizes exposure of the stack and BoP components to strongly oxidizing conditions. Air is typically filtered to remove solid particles and dried to decrease its steam content., effectively mitigating oxygen electrode decomposition and enhanced Cr evaporation from interconnects and BoP components. Nevertheless, the necessity of air drying under all ambient conditions remains uncertain. In dry regions, air drying may be unnecessary potentially reducing CAPEX, but further validation under real-world conditions is needed. In addition, air filters capable of removing sulfur impurities were shown to substantially decrease degradation.[90]

However, using air as sweep gas, can introduce nitrogen into the fuel gas in the case of internal leakages between the fuel and the oxygen electrodes, necessitating additional purification. Moreover, depending on the pressure level and pressure drops, air sweep gas requires an air compressor or blower and an electric heater, which increase CAPEX increase and reduce overall system efficiency. Consequently, a system design without air sweep gas could be economically favorable if safety issues and temperature control can be adequately addressed,[119] while also preventing impurity-related degradation and nitrogen impurities in the product gas.[120] Manufacturer guidelines for inlet stream quality are usually conservative, more research and practical experiences are required to define appropriate purity specifications.

#### *5.2.3 High-temperature heat exchangers and electric heaters*

The gaseous inlet streams are pre-heated in HT HEXs via recuperation of the respective hot outlet gases and subsequently typically heated up to operating temperature with electric trim heaters. HT HEXs, electric heaters and hotbox piping steels operate at high temperatures in challenging gas environments, facing the risk of lifetime-limiting corrosion. Upstream metallic components can also release chromium, leading to oxygen electrode poisoning, and Si, depositing in the fuel electrode, both accelerating electrode degradation.[121]

*Materials*

Cr evaporation and silica carryover are temperature-activated,[122] and as a result, advanced material selections such as high-grade ferritic and austenitic stainless steels or nickel-base alloys may be required depending on the specific SOEC operating temperature. Coatings to diminish chromium evaporation and corrosion as used in interconnects are difficult to implement for BoP components due to complex shapes. However, there are no restrictions regarding the electrical conductivity of surface oxide scales and the CTE which provides more flexibility in material selection.

The most cost-effective ferritic steels display significant corrosion on both fuel and oxygen side as well as Cr evaporation during typical SOEC operating temperatures above 600°C. [122-124] However, reducing the operating temperature to a range where mass-manufactured ferritic stainless steels can be used, as for instance in state-of-the-art MSC architectures,[38] could considerably improve business cases by reducing BoP component costs.

Austenitic steels are often favored due to their higher strength, creep resistance, higher corrosion resistance and microstructural stability.[125] Above 650°C, they show good Cr retention capabilities but also significant corrosion,[122] especially in $H_2/H_2O$ atmospheres.[126] Hence, nickel-base alloys are favored above 650°C due to superior corrosion resistance and mechanical properties, although they are 3-10 times more expensive than iron-based alloys.[127] For this reason, the cost for HT HEXs increases considerably above 600-675°C. [127-129]

FeCrAl alloys also offer high-temperature corrosion resistance due to slowly growing protective Al scales, which can significantly diminish Cr evaporation.[122] In particular, Cr and Al-lean FeCr alloys (2-6 % Al, 5-15 % Cr) have recently demonstrated excellent chromium retention and corrosion resistance in humid air and hydrogen atmospheres up to 850°C.[126, 130] Although more expensive than most ferritic and austenitic stainless steels, they may be suitable for temperatures above 650°C. However, they may lack the required creep strength for BoP applications above 800°C. For this reason, recently alumina-forming austenitic steels have been developed that combine high creep and corrosion resistance at lower cost than Ni base alloys.[131, 132]

*HT HEX*

During $CO_2$ and co-electrolysis, carbon deposition can occur on the fuel side outlet of HT HEXs due to the CO-rich gas atmosphere at 450-750°C, in particular at elevated pressures.[133] This may reduce heat transfer, increase pressure losses or cause full blockage. Furthermore, metal dusting can lead to HEX damage and eventually to premature failure. Common strategies to avoid carbon deposition/metal dusting focus on using materials that can either suppress catalytic CO decomposition or prevent carbon adsorption.[134] For instance, alloys with Al and Si can form protective oxide surface layers.[134] However, complete avoidance of coking is unlikely. Therefore, occasional HEX regeneration via oxygen-containing purge gases, transient temperature increases or soot blowing were proposed to prolong lifetimes.[135] Introducing small amounts of steam into the exhaust gas can also mitigate coking by increasing the $pO_2$, but comes with efficiency penalties. Alternatively, designing the HT HEX for rapid temperature transitions through the critical 500–700 °C range, where carbon formation is thermodynamically and kinetically favorable, can minimize residence time and carbon formation. Mechanically, large tube diameters reduce pressure drops and the need for high-power blowers, but increase the HEX surface area and overall volume.

*Electric heaters*

During operation, after preheating the inlet gases in the HT HEX, additional electric trim heaters are typically used to raise the gas temperatures to the target level. This is required even at isothermal operation to compensate thermal losses, and offset mass losses on the fuel side caused by oxygen ion migration to the oxygen side. Furthermore, electric heaters, in particular on the oxygen side, are essential for providing the required heat during start-up. Few electric heaters for HT use above 700°C are commercially available,[136] generally based on FeCrAl alloys, which display low Cr evaporation and corrosion rates on the air side at temperatures up to 850°C.[137, 138] However, dedicated long-term investigations of electric heater stability and potential contaminant release are lacking. If heater lifetimes are shorter than those of the stacks, exothermic stack operation reduce heater demand and increase their lifetime. Moreover, heating via inlet gases is inherently inefficient, as the outlet streams inevitably retain significant residual heat, leading to substantial thermal losses.[139] Alternative concepts using a furnace around the stack module, external heating plates or radiant heaters integrated into the hotbox have been proposed to reduce energy consumption and avoid contamination release, potentially reducing energy consumption by up to 50%.[140, 141] Such an approach could also enable operation without air sweep gas.

### *5.2.4 Low-temperature heat exchangers/economizers*

Since the cold fuel gas feed enters the HT HEX at temperatures above 100°C, the hot fuel gas exits at an equivalent or higher temperature, leaving a significant amount of unrecovered heat, including the latent heat of steam evaporation. Recovering this remaining off-gas heat is vital for achieving high system efficiencies in use cases without substantial external heat sources.[142] To maximize heat recovery in highly integrated systems, the liquid feed water is used to absorb the residual heat from the off-gas in a LT HEX, also referred to as an economizer.[142-144] The economizer is often integrated into a HRSG since sufficient off-gas heat is available to partially evaporate the feed water, but can also be implemented as a separate standalone LT HEX upstream of the steam generator. In contrast to the HT HEX, off-the-shelf products for low-temperature applications constructed from materials such as steel, copper or aluminum, can be used. However, due to their lower $\Delta T$, LT HEXs must be larger to achieve sufficient heat transfer, which increases costs, and the off-gas still exits above 25 °C, containing 5–10 % residual steam whose latent heat cannot easily be recovered.

Alternatively, if sufficient waste heat is available on-site, the LT HEX can be replaced by a condenser with chilled cooling water which can further reduce the steam content in the off-gas well below 1 % but does not recover the remaining heat.

On the oxygen side, excess heat is frequently available after the HT HEX due to the higher mass flow at the stack outlet. This residual heat can be used for water pre-heating in an additional LT HEX,[145] or for recirculation stream heating. However, the amount of excess heat depends strongly on the generated oxygen production rate which is proportional to the operating current density.

*5.2.5 Recirculation*

Recirculation of part of the product gas into the fuel gas inlet stream is required to maintain the state-of-the-art Ni/cermet electrodes in a reduced state. Three main recirculation strategies are generally considered (Figure 8): 1) HT recirculation, 2) intermediate-temperature (IT) recirculation after the HT HEX, and 3) recirculation after the LT HEX or condenser. In all cases, recirculation blowers are required to compensate the pressure losses of the stack module. At fixed single pass conversion, both HT and IT recirculation enable higher system RCs since the steam in the off-gas can be re-utilized (Figure 7), which reduces energy consumption for steam generation.

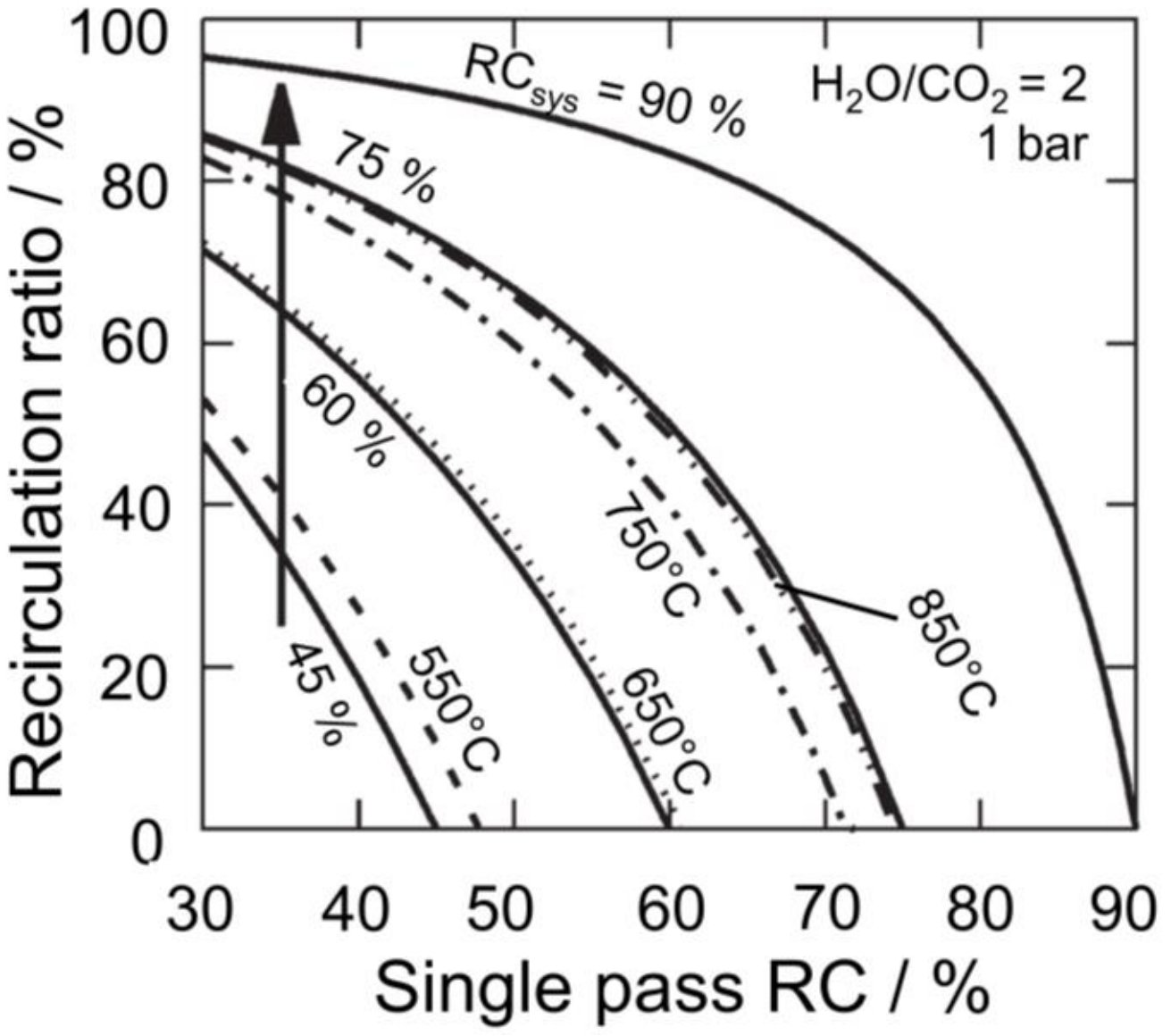

Figure 7. System RC dependency (solid lines: 45, 60, 75, 90 %) on single pass RC and recirculation ratio at $H_2O/CO_2$ inlet ratio of 2. Dashed lines show thermodynamic tendency of carbon formation at different temperatures (550, 650, 750, 850°C). Modified with permission

based on Gas Recirculation at the Hydrogen Electrode of Solid Oxide Fuel Cell and Solid Oxide Electrolysis Cell Systems, Henke et al., Copyright 2016, Wiley.[146]

HT fuel gas recirculation (Figure 8a) is attractive, since the recirculated gas does not require re-heating, potentially decreasing HT HEX size and electric heater power consumption. Hence, this approach is often used in concepting studies.[147, 148] However, the development of reliable high-temperature blowers is technically challenging and no mass-market products are currently available.[146, 149] Furthermore, flow rates of the recirculated off-gas cannot be determined directly due to temperature restrictions of the sensors and must instead be estimated from characteristic blower map as a function of rotational speed, temperature and pressure.

Alternatively, ejectors made from high-temperature steels do not contain any moving parts and are for instance, used in the 250 kW SOEC system by Convion Ltd. (Finland).[150] Their main drawback is the fixed recirculation ratio which is dictated by the ejector geometry. This presents a pressure management challenge in maintaining stable flow and pressure across the SOEC stack, especially during load changes. One mitigation strategy involves supplying a supplementary feedstock stream that dynamically adjusts the recirculation ratio for stable operation and pressure control, and is cut off at nominal operation.[151]

Alternatively, the product gas can be cooled via the HT HEX and subsequently recirculated at intermediate temperatures of ~200-300°C (Figure 8b).[152] At these temperatures, the technical feasibility and reliability of fuel off-gas recirculation blowers are significantly improved. For instance, AVL List GmbH (Austria) has developed an SOFC off-gas blower operating at up to 600°C.[153] Furthermore, recirculation flow rates can be determined with higher accuracy due to the availability of measurement devices at these temperatures allowing improved control of the recirculation process. However, in this approach the recirculation pipes are located outside the hotbox, necessitating additional insulation, and the size of the HT HEX may be larger due to the higher heat transfer duty.

After cooling in the LT HEX (Figure 8c), only a small amount of steam remains in the fuel gas (for instance, 7% $H_2O$ at 40°C outlet temperature). As a result, system RC can barely be increased by recirculation, leading to a higher specific energy demand of the evaporator.[106] On the other hand, the low gas temperatures in the recirculation loop minimize the need for pipe heating to prevent condensation and the blower energy consumption is reduced, since most of the steam has been condensed before. For LT blowers, synergies with existing PEM fuel cell recirculation blowers could be leveraged, potentially reducing CAPEX.

Ceres has demonstrated that their MSC stacks with Ni/CGO fuel electrodes can operate at 95% RC in steam electrolysis with 100% steam feed, indicating the feasibility of simpler systems without recirculation loops.[154] However, long-term validation is still required to evaluate possible degradation due to partial Ni oxidation at the fuel inlet or the fuel electrode/electrolyte interface.

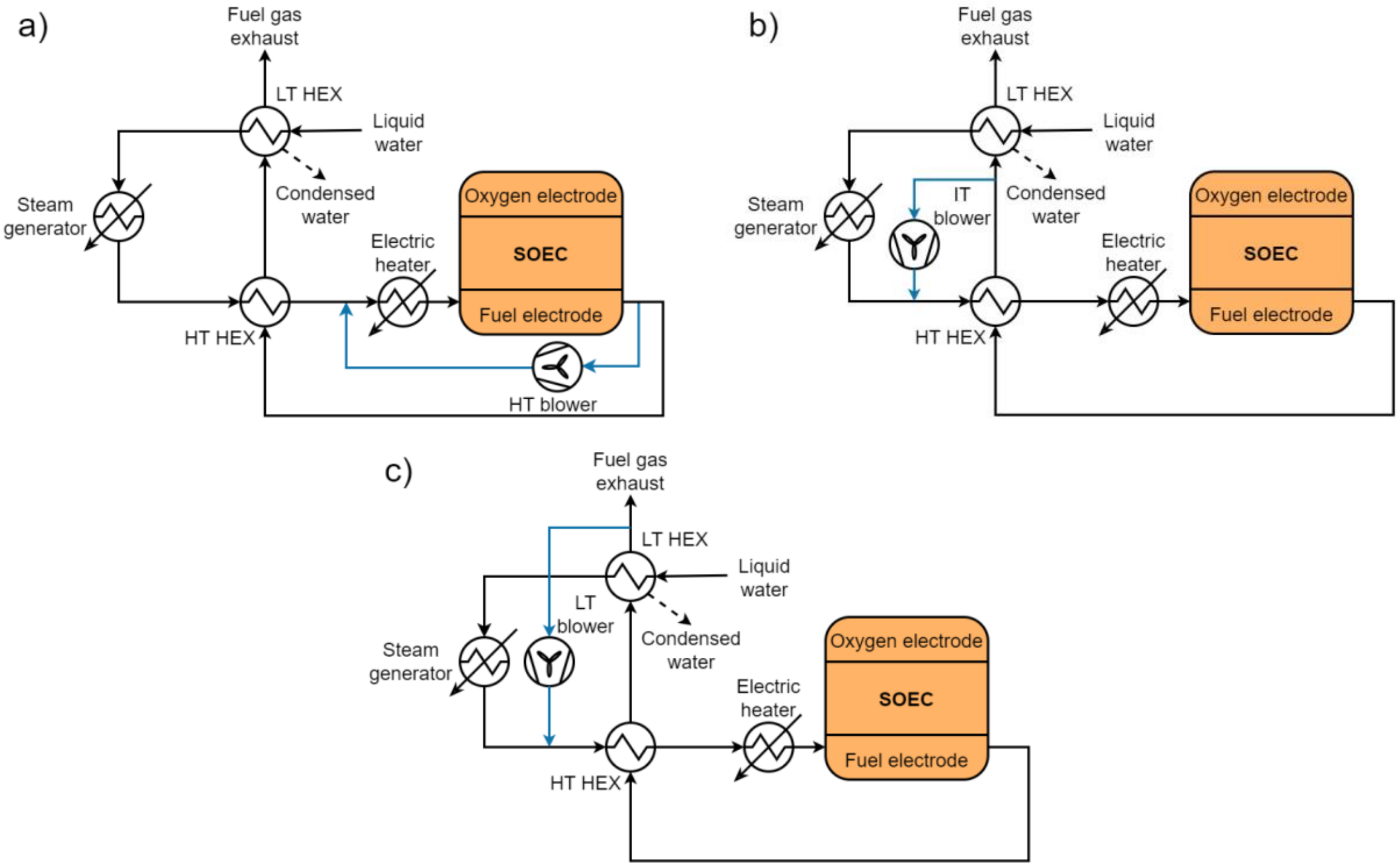

Figure 8. Simplified process flow diagrams for three different fuel off-gas recirculation strategies: a) high-temperature (HT) recirculation directly after the exhaust, b) intermediate-temperature (IT) recirculation after the HT HEX, and c) low-temperature (LT) recirculation after the LT HEX. Light blue lines indicate the recirculation lines. The streams on the oxygen side are omitted to enhance visibility.

### *5.2.6 Downstream product treatment*

Before further processing, the produced hydrogen or syngas is primarily purified water removal. Drying is typically performed to meet the required purity specifications using pressure swing adsorption (PSA) or temperature swing adsorption (TSA) units.[155] TSA is usually favored due to a lower product gas losses during the desorption phase. Zeolite molecular sieves or silica gel are frequently used as adsorbents for water removal, while additional adsorbents such as carbon molecular sieves or certain zeolites can also be incorporated in multi-stage TSA/PSA units for comprehensive impurity or inert gas removal. Achieving high product gas purity necessitates elevated operating pressures, [156] for instance, 10-30 bar to reach hydrogen 5.0 specifications with <5 ppm $H_2O$. Consequently, multi-stage compression of hydrogen/syngas is required upstream, which significantly increases the gas temperature. To improve compression efficiency and prevent overheating of the commonly used membrane or dry running piston compressors above 150-250°C,[155] intercooling is carried out between stages. Part of the intercooling waste heat can be recovered for SOEC feed water pre-heating or steam generation. Water condensation is enhanced at elevated pressures due to increased water vapor partial pressure, enabling efficient removal at the intercooling stages.

Although DeOxo reactors are generally unnecessary in SOEC systems, since any crossover oxygen reacts spontaneously with hydrogen at temperatures above its autoignition point (~560 °C), they may still be installed to ensure ultra-high purity.

For storage and transportation, further compression of hydrogen is required, and downstream synthesis reactors are operated at pressures up to 200 bar. Due to high specific heat of hydrogen, compression of hydrogen-rich gases requires more energy than other gases, such as natural gas.[157] Hydrogen's small molecular size also imposes strict material requirements to prevent permeation and embrittlement, contributing to high compressor costs. Safety is further complicated by the Joule-Thomson effect: Hydrogen heats up upon expanding at temperatures above its inversion point (-80°C), increasing reactivity which becomes a risk in case of a leakage.

*5.2.7 Power electronics*

Alternating current (AC)/direct current (DC) converters are the largest contributors to CAPEX alongside the stacks in SOEC systems.[8, 158] Therefore, large-scale adoption of SOEC requires the availability of power electronics in adequate sizes and quantities. A typical SOEC stac requires high currents (e.g., 240 A for a stack with 300 $cm^2$ active area at 0.8 A $cm^{-2}$) and high voltage (e.g. 150 V for 100 cells at 1.5 V/cell). State-of-the-art power electronics devices for such applications include thyristor-based rectifiers and chopper-rectifiers using insulated gate bipolar transistors (IGBTs).[159] Thyristors-based rectifiers deliver high power reliably at high efficiencies ~ 97%,[160] but produce significant harmonic distortions and reactive power especially during part-load operation. This necessitates the use of large active and passive filters to improve power quality which increase CAPEX and size. Diode rectifiers with DC choppers show higher power quality with reduced ripple current, but are more complex and less reliable than thyristor-based rectifiers.

Reversible solid oxide cell operation is carried out with bidirectional power electronics which show substantially lower efficiencies due to their wide operating range. Separate unidirectional power electronics have been suggested to increase efficiencies.[161]

High current ripple was shown to increase losses in AEL, negatively affecting specific energy consumption.[162] However, the impact of current ripple on SOEC performance and long-term stability remains uncertain and the demonstrated stack ability to withstand a certain degree of ripple current would enable simpler and more cost-effective power electronics designs.

### *5.3 Pressurized operation*

From a system's perspective, pressurized steam electrolysis operation has three main advantages. First, pressurization of incompressible liquid water upstream of the SOEC reactor is less energy-consuming than the energy-intensive hydrogen or syngas compression downstream. Process simulations have suggested pressurized steam electrolysis operation to potentially increase system efficiency by 3-7 %.[56, 163] Second, at elevated pressures heat exchanger areas required for effective recuperation are decreased which can reduce system CAPEX.[164] Third, for atmospheric SOEC operation, the first product compression stage is particularly costly due to its relatively large size. Pressurized SOEC operation can eliminate the need for this initial compression stage already at lower operating pressure around 2-3 bar,[155] resulting in substantial cost savings.

In most experimental demonstrations the entire cells or stacks were integrated into a pressure vessel to avoid high differential pressures between electrodes and from the electrodes to the environment.[56, 164-166] In this way, pressurized SOEC operation at up to 25 bar has been demonstrated on the stack and system level.[56, 165, 167, 168] Since in this approach the oxygen side inlet gas is compressed as well, efficiency gains are maximized in systems that do not use air as sweeping gas since the energy demand for air compression may offset all potential savings on the fuel side.[119]

Operation under differential pressure can eliminate the need for air compression and increase system efficiencies. Thermodynamically, this approach is favorable since the OCV of SOEC generally increases with pressure according to the Nernst equation, and lower oxygen partial pressures can mitigate this effect.[169] Furthermore, internal leakage of nitrogen into the fuel gas is suppressed under differential pressure.[170] Stack modules developed and commercialized by Topsoe can be operated under differential pressures of up to 10 bar.[171] The stacks have shown high stability at 2.5 barg for >2000 h, and the capability to withstand 7 bar differential pressure between the electrodes for 70 h, however, long-term stability beyond 10 kh still needs to be demonstrated.

The effect of pressure on performance is architecture-dependent. For FESCs, ASR decreases significantly with pressure due to decreased electrode and diffusion resistances which can compensate the increase in voltage according to the Nernst equation.[164-166] By contrast, ESCs show no such benefit, resulting in slightly reduced current density at isothermal operation.[165, 167]

Higher operating pressures could also lead to accelerated kinetics of partial pressure-dependent degradation processes (e.g. corrosion of the interconnector on both sides,[172] Ni migration/coarsening), but only few degradation studies are available under relevant operating conditions.[173]

Additionally, in pressure vessels system-level heat management becomes increasingly challenging, since the insulation performance of typically used microporous materials decreases with pressure, leading to higher hotbox heat losses.[56]

In summary, the optimal pressure for steam electrolysis is limited by i) stack robustness and pressure-induced degradation, ii) architecture-dependent performance, and iii) system-level heat management, all of which require further investigation.

In pressurized co-electrolysis, additional challenges arise. Thermodynamics dictate an increase in methane and carbon formation according to eqs. 5, 6 & 8. The absence of hydrogen prevents methanation in pure $CO_2$ electrolysis, but further increases the risk of carbon formation. Safe operating conditions must therefore be strictly maintained to avoid carbon accumulation. Moreover, metal dusting in the fuel side heat exchanger is promoted under pressure,[174] highlighting the need for advanced material solutions.

Increased methane outlet concentrations at elevated pressure may be favorable for coupling with methanation reactors,[106] but are undesirable for methanol or FT synthesis.[119] Furthermore, elevated pressures can lead to higher stack temperatures due to accelerated reaction rates of exothermic methanation which lowers the operating voltage, but also the current density at isothermal operation.[175] As a result, pressurized co-electrolysis introduces significant operational and material constraints, and its practical benefits for specific applications remain uncertain.

### 5.4 *Pathways for syngas production*

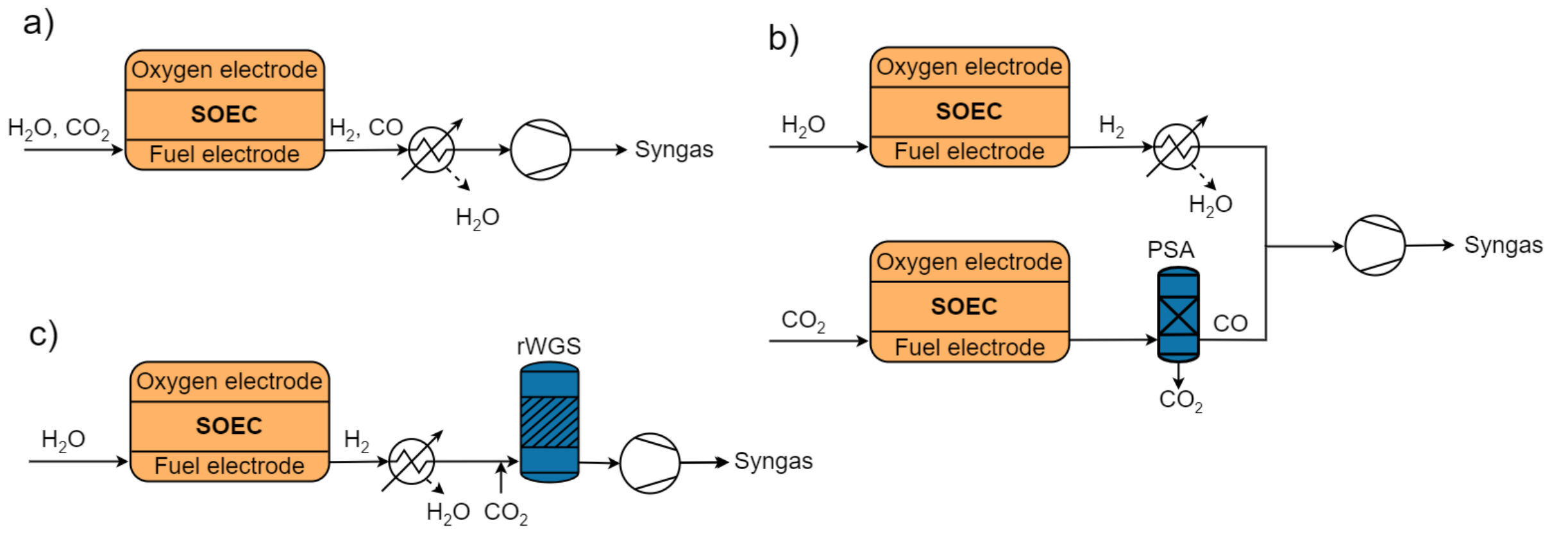

Figure 9. Schematic illustration of the three routes for syngas generation. a) Co-electrolysis, b) separate steam and $CO_2$ electrolysis, and c) steam electrolysis coupled with a RWGS reactor.

Based on solid oxide electrolysis, three process routes can be envisioned to generate syngas (Figure 9).

In the co-electrolysis route (Figure 9a), a single reactor type is used for syngas production, with the syngas ratio (SGR) depending on the operating parameters such as temperature, pressure, current density, inlet fuel gas composition and fuel gas flow rate. By tuning these conditions, most prominently the inlet gas phase composition, a high flexibility can be achieved enabling tailor-made SGRs.[176] For operating temperatures above 750°C the fuel gas outlet can be predicted by the thermodynamic equilibrium with high accuracy, and the dry syngas can still contain between 9-16% of undesired $CO_2$.[167] Co-electrolysis operation allows higher current densities at isothermal operation compared to steam electrolysis due to a higher thermoneutral voltage and the cooling effect of the endothermic RWGS reaction. For instance, a 13% higher current density was reported for ESC-based SOEC.[117]

However, the propensity for methane and carbon formation at low temperatures, high pressures and high RCs can restrict the co-electrolysis operating window. Carbon formation must be avoided due to the risk of irreversible stack damage. Exothermic methane formation leads to stack heating, which reduces the isothermal current density and causes mechanical stresses. Both challenges become particularly relevant below 700°C, which may limit the use of LT SOEC technologies such as MSCs for co-electrolysis. By contrast, SOEC operated at higher temperatures >800°C such as ESCs are less sensitive towards coke and methane formation.

The electrochemical reduction of $H_2O$ and $CO_2$ in separate reactors (Figure 9b) may increase flexibility in tailoring product gas mixtures. However, coking issues are even more severe in $CO_2$ electrolysis, necessitating operation at lower RC values that result in outlet gases with high $CO_2$ content. For instance, $CO_2$ conversion in FESCs was reported to be limited to ~45 %.[78] Depending on the downstream process, this may necessitate an additional $CO_2$ separation unit, increasing CAPEX. Furthermore, the thermoneutral voltage of ~1.47 V in $CO_2$ electrolysis is significantly higher than in steam electrolysis which can make the generally preferred isothermal operation unfeasible due to the increased risk of electrochemically induced carbon formation at high fuel electrode overpotentials, especially in FESCs.[84] As a consequence, this process route has received relatively little attention so far.

A third option combines steam electrolysis with a dedicated RWGS reactor (Figure 9c). The RWGS reactor converts $H_2$ and $CO_2$ at 700-1000°C to suppress the formation of methane and coking. RWGS reactors are still under development and have not yet been demonstrated on an industrially relevant scale yet.[177] When coupling a RWGS reactor with a steam electrolyzer, water must first be removed from the SOEC off-gas gas to allow thermodynamically more favorable RWGS conditions. Hence, cooling, water knock-out and re-heating of the product gas are required. In addition, the endothermic RWGS reactor requires additional external high-temperature heat. The added complexity, including additional heat exchangers, electrical heaters and an additional reactor, will likely increase CAPEX and decrease system efficiency compared to direct co-electrolysis. For example, syngas production via co-electrolysis was estimated to achieve a 2-7 % higher efficiency than steam electrolysis + RWGS pathway when an electrical steam generator is used, since extra steam is required to generate additional hydrogen required for $CO_2$ reduction in the RWGS reactor.[119] Moreover, co-electrolysis also requires a lower SOEC area due to the higher current densities at isothermal operation.[119]

However, the combination of steam electrolysis with RWGS may be a viable pathway for pressurized LT SOEC technologies, where direct co-electrolysis leads to a high risk of methane and carbon formation.

## 6 Operating and control strategies

Efficient and reliable operation of SOEC systems depends on effective operating and control strategies to maintain optimum performance and minimize degradation. Main control parameters include the electrical current, stack voltages and stack temperatures, that all need to be kept within defined boundaries to ensure safe and efficient operation. Characteristic temperatures are typically monitored either inside the stack or at the stack outlet. Due to the non-linear temperature behavior with current density/voltage (Figure 10), these parameters are strongly interdependent. Closely related to power control are the recirculation rate (RR) and RC, which must be regulated to avoid reactant starvation which can accelerate degradation and increase specific power consumption. Furthermore, the SGR needs to be controlled in co-electrolysis to fulfill the requirements of the downstream process.

Control strategies so far have largely focused on simple single input, single output (SISO) proportional-integral-derivative (PID) or feed-forward (FF) controllers due to high reliability and low computational effort. Model predictive control (MPC) provides a more sophisticated approach since it can handle multiple inputs and multiple outputs simultaneously, making it more suitable for complex SOEC systems with highly dynamic behavior and multiple constraints. Furthermore, MPC could reduce the number of sensors in SOEC systems for cost considerations. For example, temperatures in the entire stack could be inferred based on one characteristic temperature using a validated underlying model. However, MPC is computationally more demanding and coordination of multiple control targets can be a challenging task.

In the following, operating and control strategies for SOEC systems in stationary and dynamic operation are outlined.

## *6.1 Stationary operation*

*Operating regimes*

SOECs can be operated in endothermic, exothermic or isothermal modes, depending on the operating voltage (Figure 10). Endothermic operation allows stack efficiencies >100 % according to eq. 7, but requires a rarely available external high-temperature heat source to heat the inlet gases to the operating temperature. Supplying the heat electrically reduces overall system efficiency,[178] and lower operating current densities increase the number of stacks needed, increasing CAPEX. Exothermic operation avoids the need for external heating but results in stack efficiencies below 100 %. Both endothermic and exothermic operating modes generate thermal gradients in the stack.

Instead, continuous, isothermal operation is generally preferred since it minimizes temperature gradients and associated mechanical stresses. In practice, the isothermal operating voltage is slightly above the thermoneutral voltage to compensate heat losses to the environment (Figure 10). In controlled furnace environments, the core stack temperature is same at OCV and isothermal operation, and is used to determine the isothermal operating point. In actual system environments with less uniform temperature distribution and fewer sensors, isothermal operation is defined by equal inlet and outlet reactor temperatures. A wide range of isothermal operating points can be achieved through combinations of current density, stack temperature and feed gas flow/composition. Operating temperatures can be adjusted by multiple operating parameters, such as electrical heater power on fuel and oxygen side, fuel and air flow rates and RRs.[53] This enables isothermal operation at part load down to 5% of nominal load.[179] However, some systems, such as the Sunfire 10.7 MW system, operate only down to 50% load, possibly due to the absence of an air purge gas stream for temperature regulation.[180] Ultimately, the nominal operating point is chosen as a compromise between hydrogen/syngas production rate and durability.

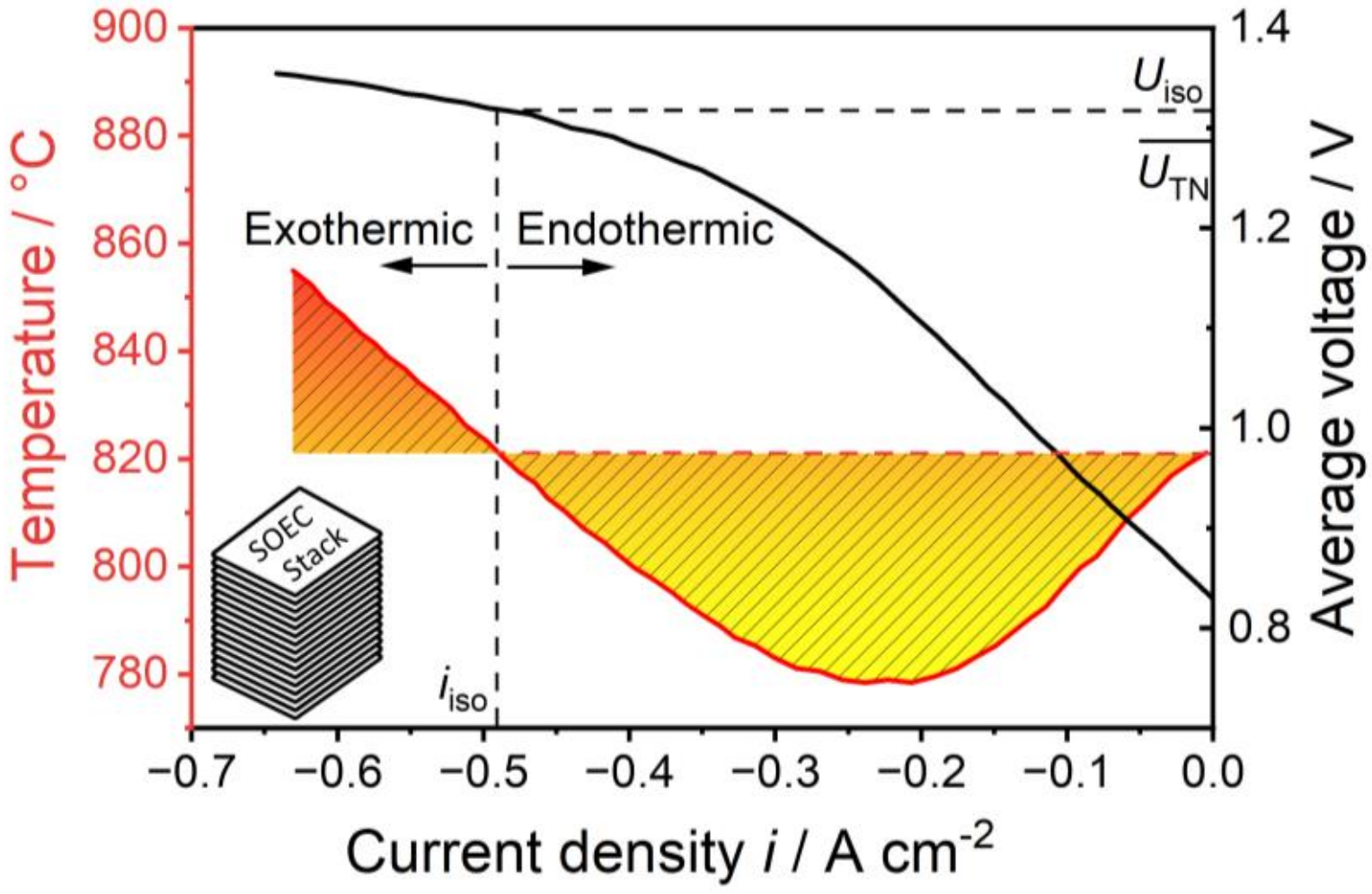

Figure 10. Exemplary current-voltage characteristics and core temperature of a stack based on 30 ESC in steam electrolysis. Image was generated based on data taken from Ref.[63]. $U_{iso}$ and $U_{TN}$ denote the isothermal and the thermoneutral voltage, respectively.

*Compensating degradation*

Current-controlled (galvanostatic) SOEC operation enables control of production rates which facilitates integration with downstream processes. However, maintaining a fixed current leads to gradually increasing overvoltages due to stack degradation, which induces thermal gradients in the stack. In contrast, voltage-controlled (potentiostatic) operation maintains the stack voltage within safe operating limits and supports thermally balanced operation by mitigating such gradients. However, as degradation progresses, the declining SOEC current and production rates cause the downstream reactors, initially sized for nominal output, to become increasingly oversized.

As a countermeasure, degradation effects can be compensated by gradually increasing the operating temperature.[58, 181] This operating strategy requires system designs that accommodate broader temperature windows, which may increase CAPEX. Degradation is accordingly reflected in the rate of temperature increase rather than in the ASR increase, which must be determined via temperature correction to the initial reference point. Furthermore, stack temperatures cannot be increased indefinitely since they are constrained by material limitations such as the maximum temperatures of glass sealant in ESCs/FESCs or CGO-based electrolytes in MSCs. Consequently, switching from full load to part load once the maximum temperature has been reached can extend stack lifetime. Determination of the operating strategy to maximize the overall stack lifetime capacity in terms of cumulative hydrogen/syngas production while minimizing costs requires economic evaluation.[182]

The outlined approaches allow SOECs to maintain near-constant stack and system efficiency over time, unlike low-temperature electrolyzers, which are typically operated galvanostatically at nominal operating temperature and become less efficient over time due to increasing heat losses. For low-temperature electrolysis, stack replacements have been suggested to be economically viable after a ~15 % voltage increase, corresponding to a similar end-of-life (EoL) reduction in efficiency.[107]

*Reactant conversion control, recirculation strategies and syngas ratio*

Typical single pass RC values for steam electrolysis are between 70-85 %. In SOEC stack modules, inhomogeneous fuel flow distribution between the stacks and potential leakages could lead to locally increased RCs and an increased degradation risk. Such inhomogeneities can be detected using temperature and voltage sensors. System RC values can be increased via HT or IT recirculation, which reduces specific evaporator power demand and increase system efficiency in case of limited availability of excess heat for water evaporation.[106] System RC values can be increased in particular at low single pass RC, whereas the benefit is smaller if single pass conversion is high.[183] However, to obtain high system RC values, relatively high RRs are required. For example, a single pass RC of 70 % may require a RR of 74 % to achieve a system RC of 90 % (Figure 7), independent of current density and stack technology.[146] For such high RRs, the increase of the Nernst potential due to increased reactant concentrations will lead to lower operating current densities. High RRs also increase pressure drops, in turn leading to increased blower power consumption which can decrease system efficiency.[178]

$CO_2$ and co-electrolysis operation may necessitate considerably lower RC system values due to the risk of carbon formation, which can be assessed by the O/C ratio. Since thermodynamic carbon formation is thermodynamically most favorable at the fuel gas outlet, effective system RC control enables the adjustment of the O/C ratio to maintain safe operating conditions. The maximum system RC typically decreases with decreasing temperature and increasing pressure,[146] and the risk may be further increased in FESCs with thick fuel electrode support layers, especially at high fuel electrode overpotentials.[78] More work is required to determine safe operating conditions that can prevent carbon formation. Since the risk of carbon formation becomes high for system RC values above 75% for all operating conditions (Figure 7),[146] and such values are frequently already achievable in a single pass, RRs during $CO_2$ and co-electrolysis are likely limited to obtain the minimum concentration of reducing gas at the inlet. A value of 10% reducing gas is frequently targeted, but lower values may be possible. The reducing gas is usually assumed to be the sum of $H_2$ and CO, with the potential addition of $CH_4$ if co-electrolysis operating conditions are favorable for internal methanation. The internal reforming of methane at the inlet can lead to localized cooling and an effective increase in reducing gas, but the effect of methane recirculation on thermal stresses and reducing gas thresholds is unclear. Generally, more research is required to find system configuration with optimized fuel gas recirculation strategies.

During co-electrolysis, the SGR must be held within a specific target window to meet downstream reactor requirements. SOEC temperature increase to counteract degradation at constant RC may result in small changes of the SGR over time. These can be corrected by adjusting the inlet $H_2O/CO_2$ ratio based on stack temperature measurements combined with thermal equilibrium calculations or look-up tables, or via online gas analysis upstream of the synthesis reactor.

### *6.2 Dynamic operation*

Due to the mechanical fragility of SOC, temperature changes are ideally performed as slow as possible. This includes controlled heating and cooling of the stack, as well as unintended thermal and load cycling due to equipment failure. Dynamic operation in fuel cell and electrolysis mode are typically avoided. However, more dynamic operation can be expected to become of higher importance in future energy scenarios when electrolysis systems will be coupled to intermittently available renewable energy sources (RES). Such coupling would likely decrease capacity factors and require operating strategies for periods without RES availability. Dynamic operation is essential in off-grid systems, for instance, wind in off-shore systems can fluctuate by more than 30 %.[184] Moreover, dynamic operation may also be advantageous in grid services applications depending on the availability of cheap electrical energy. In this regard, electricity grid operators in Europe demand response times for primary regulation of less than 30 s (or frequency containment reserve (FCR)), for secondary regulation less than 5 mins (or automatic frequency restoration reserve (aFRR)) and for tertiary regulation less than 15 mins (or manual frequency restoration reserve (mFRR)).

*Start-up and shutdown*

A cold standby/shutdown would be preferred in cases of seasonal RES unavailability where SOEC stacks are cooled down to ambient temperature with protective gases, and subsequently no gas supply and power for the electrical heaters are required to keep the SOEC at high temperature. However, re-heating (and also cooling) is usually carried out with ramp rates of ~1-2 K $min^{-1}$ and it can take more than one day until the entire system is brought to the operating temperature due to the large thermal mass of the system. For instance, the 720 kW SOEC system installed by Sunfire in Salzgitter required 36 h start-up time from cold standby.[185]

*Hot standby*

Thus, during the short-term absence of RES, hot standby is preferable. In this idle mode, electrical heaters maintain the SOEC close to the operating temperature.[117] In hot standby, the inlet temperature is usually larger than the outlet temperature due to the heat losses of the SOC. The reducing gas flow is as low as possible and continually recirculated with small purging rates depending on the leakages within the SOEC. Thus, the heat intake via the fuel gas stream has only a small effect on the reactor temperature. Instead, air inlet temperature and flow rate are used to control the SOEC temperature. The higher the SOC temperature is maintained during hot standby, the more electricity will be consumed for the electrical heaters and air compressors.[117] However, larger SOEC modules show lower heat losses due to improved surface-to-volume ratios although this can also result in longer cooling times. In their 350 kW SOEC system, Topsoe has reported a hot standby electricity consumption of 3% compared to full load operation.[186] In their 10.7 MW system, Sunfire reports a hot standby electricity consumption of 1.6 % compared to full load.[180] Temperatures below the nominal operating temperature could help to increase power consumption but will increase transition time to operation. More research is required to help identifying suitable hot standby conditions for different SOC technologies depending on the expected SOEC downtime. Topsoe and Bloom Energy specify a hot start-up time of 10 min for their systems,[104, 187] which is compatible with mFRR response requirements and common 15-minute spot market intervals.

*Load following and reversible operation*

Direct SOEC stack coupling to RES and reversible operation with constant BoP control parameters can lead to considerably increased degradation under certain conditions,[40, 63] whereas other studies have demonstrated that stack stability under transient conditions is significantly higher than generally assumed.[53, 56, 63, 188-190] However, long-term degradation studies of stacks under transient operation are scarce and no definitive guideline values exist for temporal and spatial temperature gradients due to the stochastic defect distribution and time-dependent mechanical degradation. Sunfire, Topsoe and Bloom Energy all report nominal system ramp rates of 10 %/min,[104, 180, 187] which are often limited by the BoP components (e.g. steam generator, heaters) rather than the thermal-chemical stack behaviour.

In contrast to SOFC operation, where exothermicity increases with current density, the non-linear temperature behavior with increasing current density in the endothermic region can be exploited to facilitate transient SOEC operation. Since the SOEC temperatures at thermoneutral operation are close to those at OCV, control strategies can aim to minimize the thermal gradients during crossing of the endothermic region.[117, 191] A simple increase in linear ramp speed can lead to a faster temperature equilibration and lower deviation from the initial/final temperature, but increase peak gradients.[117] A direct step to the isothermal current density would be desirable to avoid cooling of the SOEC almost entirely. An ESC-based stack with 30 RUs from Sunfire in a furnace environment was subjected to 16,000 on/off cycles every 2 mins between the near-isothermal operation and OCV without increased degradation.[192]

However, in a hotbox environment, a temperature decrease is observed along the flow direction of the dominant heating fluid (in most cases the air flow) in hot standby, and a more homogeneous temperature distribution is formed at isothermal operation. Hence, during the transition from hot standby to isothermal operation, some stack module regions may heat rapidly during a step-change resulting high gradients. FF controllers have been proposed to limit these gradients that employ fast current ramps at low current densities and lower ramps near the isothermal point.[117]

Another operating strategy to maintain a high SOC capacity factor even during electricity off-times could be the reversed operation in fuel cell mode. Since SOFC operation is exothermic, it can sustain the reactor temperature without electrical heaters, reducing auxiliary power consumption. Reversible operation is especially attractive for energy storage, grid stabilization, or island grid applications. However, systems specifically designed for reversible operation are more complex as BoP components must accommodate differing requirements for SOEC and SOFC modes.[193] Nevertheless, even systems primarily designed for SOEC operation could be temporarily run at low loads in fuel cell mode during electricity off-times to maintain operating temperature, thereby reducing electric heater power consumption in hot standby.

*Pulse-width modulation (PWM)*

Recently, novel operating methods involving PWM have been demonstrated in SOECs to facilitate load following.[194, 195] In PWM, the current input signal is represented as a rectangular wave with varying duty cycle (fraction of time the signal is ‘on’) and high switching frequency.

This operating strategy can then be used to modulate the entire part load range between OCV and thermoneutral operation by adapting the duty cycle.

PWM can help to decouple stack temperature and hydrogen production rate and mitigate the effects of mass transport limitation.[194] The operating strategy has the potential to significantly improve SOEC ramping capabilities opening up the application in grid services and maintain operation within a safe operating window. For instance, during $CO_2$ and co-electrolysis, PWM can prevent strong cooling during part-load operation in the endothermic region, where the risk of thermodynamically favored coke formation is particularly high.[195] PWM operation can support near-isothermal stack conditions during transients under constant feed conditions and in principle enable a multifold increase in ramp rates.

A similar approach is followed in the AC:DC operating technique patented by Dynelectro (Denmark) that is using rapid reverse pulsing in the ms range (e.g. 33 Hz[196]) between endothermic electrolysis and exothermic fuel cell mode.[196, 197] With the right operating parameter choice, net electrolysis operation can be achieved while keeping the stack module isothermal (Figure 11d). Furthermore, AC:DC operation has been shown to significantly mitigate degradation mechanisms such as impurity poisoning and nickel migration, with a 4-5x increase in lifetime estimated compared to conventional DC operation.[66, 196] Nevertheless, the benefits and limitations of this operating approach require more investigations including adequate tuning of the cycling parameters and profiles. AC:DC operation also necessitates the development of cost-effective bidirectional power electronics. [198] To avoid injecting large amounts of fuel cell power into the grid, which could have unforeseen effects on grid harmonics at large scale, modular systems are envisioned to switch operation sequentially. In this setup, power generated in fuel cell mode is used to drive electrolysis in another module. While such an approach decreases the system's capacity factor and requires oversizing compared to steady-state operation, the associated costs are expected to be offset by the increased lifetime.

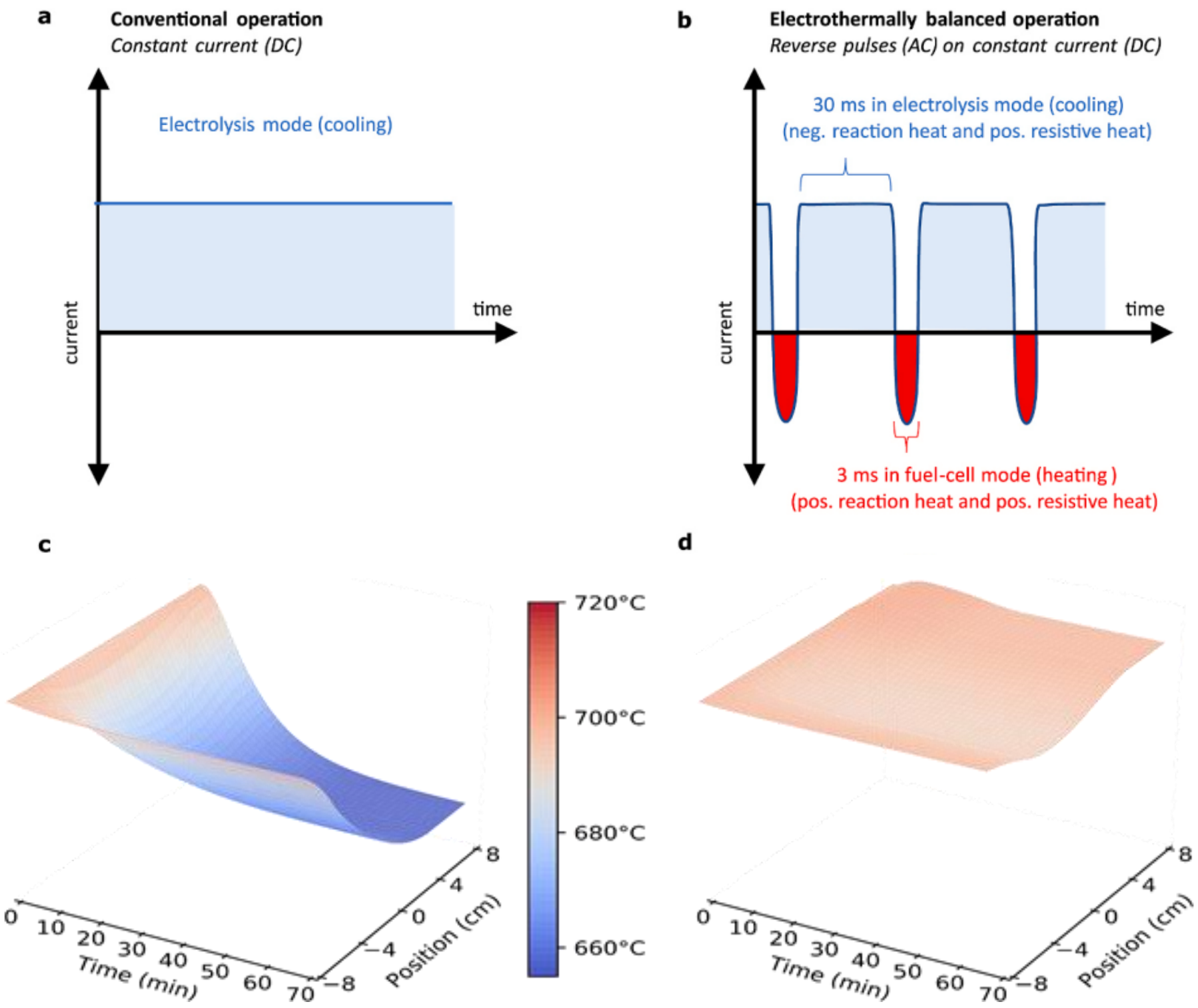

Figure 11. a) Conventional endothermic operation, (b) AC:DC operation in the endothermic operating regime with short fuel cell pulses in the ms range. (c)–(d), COMSOL-modeled temperature profiles across a stack operating in (c) DC mode, (d) and AC:DC mode. Reprinted with permission from Electrothermally balanced operation of solid oxide electrolysis, Skafte et al., Copyright 2022, Elsevier.[196]

*Operating parameter control for operational flexibility*

Furthermore, SOEC operating windows can be significantly extended by elaborate operating strategies on the system level. The fuel/air inlet temperatures, air flow rate and speed of current ramping have all been demonstrated as viable adjustment screws to minimize spatial and temporal temperature gradients.[116-118, 199, 200] For example, an air flow rate that is dependent on the load profile was shown to mitigate temperature gradients,[116, 118] whereas excessive air consumption needs to be avoided in order to maintain high system efficiencies.[201] In this regard, MPC could hold great potential to mitigate temperature gradients by varying multiple parameters simultaneously. While MPC has been explored for SOFC systems,[202] so far no work has shown its application for SOEC.

*Part-load operation with modular approach*

In addition, the modular nature of large SOEC plants will likely reduce the necessity for direct load following on stack level, enabling instead a modular on/off approach where a share of the stacks is operated at isothermal full load and a share in hot standby, supported by a battery as electricity buffer.[117] With such an approach, temporal temperature gradients can be expected to be small and transitions can be completed relatively fast. Furthermore, an extension of the operating window around the nominal full load point on the modular level can translate into a great flexibility on plant level as depicted in Figure 12.[203] While generally applicable modular operating strategies have been proposed in literature,[204] further research is required to account for the real-life electrochemical and thermal behavior of SOEC stack modules.

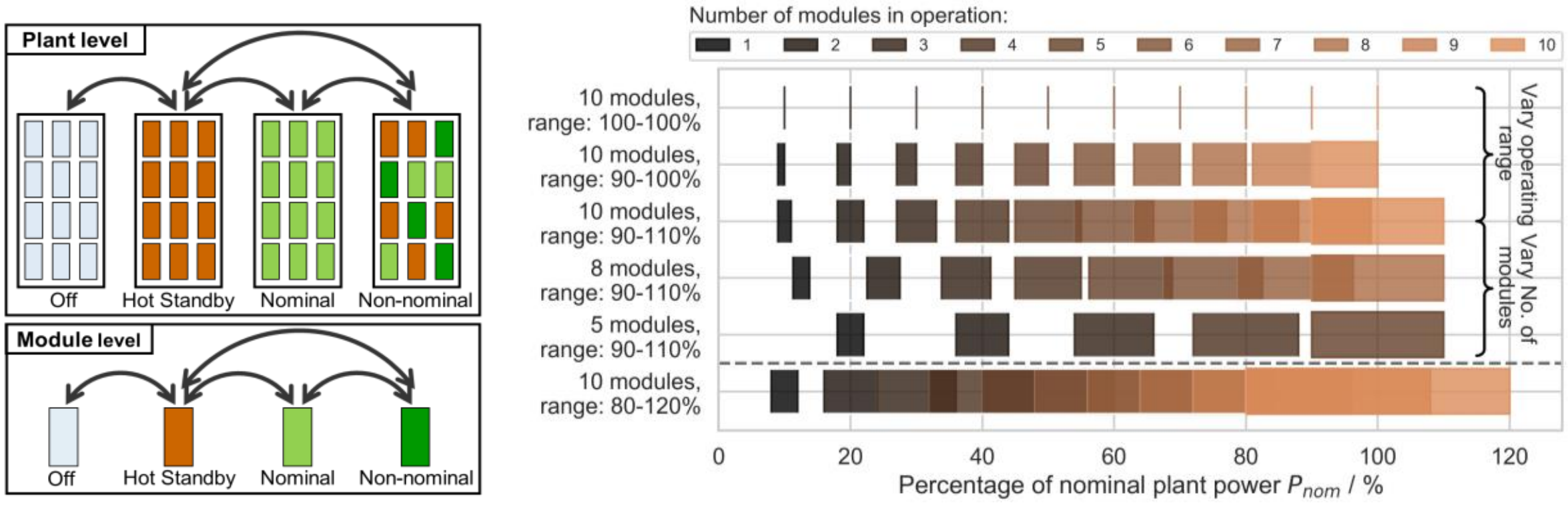

Figure 12. Accessible power range for an entire modular solid oxide electrolysis plant depending on the number of stack modules and their respective operational range described as minimum and maximum allowed power $P_{min} – P_{max}$. Modified with permission from Ref.[203].

*RC & RR control*

Since steam generation consumes a significant extent of a SOEC system's energy,[205] nominal RC should also be maintained during transient operation to reduce external heat demand.[118] RC can be controlled effectively by changing fuel inlet flow rates with carefully tuned PI controllers that avoid overshooting.[206] However, accurate control requires knowledge of both inlet and recirculated fuel flow rates and their gas composition—parameters that are difficult to measure reliably.[207] For example, recirculation gas streams are typically variable mixtures and their flow rates are challenging to quantify with high accuracy. Furthermore, online gas concentration measurements are too time-consuming for dynamic operation and associated with high costs. Model-based open-loop strategies are frequently based on simplifications and not generally valid. For SOFC systems, automotive lambda sensors have been proposed to be capable of rapidly measuring the $pO_2$ in the fuel outlet gas.[207] The resulting difference in oxygen partial pressure between the fuel inlet and outlet could potentially be used to determine the RC. Due to their low cost, these sensors could even be used in multiple stacks within a stack module to detect potential non-uniform gas distribution. Nevertheless, further development of robust RC and RR control strategies is needed, especially in dynamically operated systems.

*Switching strategies to co-electrolysis*

Temperature changes in SOEC during dynamic operation due to current density changes may lead to changes in the thermodynamic equilibrium and a fluctuating SGR. So far, limited research has been conducted on strategies to maintain a constant SGR during transient operation. To minimize the time periods of SGR deviation from the target value during start-up, a switching strategy from $H_2O$ electrolysis to $CO_2$ electrolysis under load may be more effective than switching gases at OCV. However, the start-up of co-electrolysis requires a fuel gas change from a $H_2O/H_2$ to a $CO_2/H_2O/H_2$ mixture and cause the RWGS reaction to occur in the fuel electrode leading to cooling of the SOC reactor and associated thermal gradients. To counteract the cooling and keep the reactor isothermal during the transition an increase in current density via simple PI or FF controllers has been demonstrated.[117]

# 7 Techno-economic analysis (TEA) of SOEC systems

## *7.1 Current CAPEX*

High system CAPEX remains one of the primary challenges of the SOEC technology.[158, 208] Reported system CAPEX typically includes stacks, as well as HT and LT BoP components shown in Figure 4, and is the most commonly used metric in TEA studies, including by European funding agencies and policy bodies. In addition, total hydrogen production plant CAPEX also covers buildings, electrical grid connections including utilities, switchgear, compression, indirects, contingencies, and water treatment.

Assessing SOEC system CAPEX range is difficult since little data of the largest manufacturers is publicly available. The most commonly reported system CAPEX values are in the range of 2000 to 2500 €/kW,[7, 56, 62, 209] though these values are mostly from around 2020. More recent data from Ceres suggest a significantly lower value of ~1450 €/kW for a first-of-a-kind (FOAK) SOEC system (1900 £/$kg_{H2}$/day, converted using average 2024 annual exchange rates).[105] Bloom Energy has even reported slightly lower values of 1,000-1,200 €/kW in 2023 (converted from 1100-1300 USD/kW using average 2023 annual exchange rate).[210] Nevertheless, they are significantly higher than the estimated 600 €/kW for AEL and 900 €/kW for PEMEL systems.[62]

CAPEX is most commonly reported in €/kW, but this metric can be misleading as it neglects system efficiency and hydrogen output. When expressed in €/kg $H_2$/day, SOEC systems appear more competitive due to their higher hydrogen yield per unit power input.[56]

The current initial cost share of the SOEC stacks has been estimated to be only 30 %, significantly lower than for AEL (50 %) and PEMEL (60 %).[7] Other major cost contributions to system CAPEX include power electronics (30 %) and other BoP components (34 %).
Furthermore, external infrastructure can represent a substantial share of total project cost.[211] The lower electricity demand of SOECs can reduce substation and grid connection requirements compared with low-temperature electrolysis, potentially lowering total hydrogen plant costs. However, this advantage is typically not captured in reported system CAPEX figures. Water purification units are usually excluded as well, although their cost is generally negligible for large-scale projects.[209]

### *7.2 Future CAPEX*

Estimating future SOEC costs is challenging since the technology is only now approaching large-scale commercial deployment and associated learning rates are subject to large uncertainty. Nevertheless, SOEC system CAPEX is widely expected to decrease over time as a result of technological progress, extended lifetimes, scale-up and mass manufacturing. The European Union has set an ambitious target of 520 €/kW by 2030.[62] While such projections remain highly uncertain and some TEA studies do not predict such a CAPEX decline until 2050.[7, 212]

SOEC stacks should fit the simple, standardized and modular type 1 technologies, which are expected to follow Wright's law and achieve high direct learning rates of 10-20 % through mass manufacturing production, material optimization, and assembly experience, and horizontal scaling via replication of standardized units.[211, 213] Compared to PEMEL and AEL, SOEC stack learning rates should be higher due to their comparatively low TRL but may reduce once the technology matures.

In contrast, many BoP components are mature technologies that have already come down their learning curves. Further cost decreases will likely arise from indirect learning, such as improvements in manufacturing processes, supply chains, and standardization.[211] As a result, SOEC system costs show a stronger dependency on system scale due to the higher cost share of peripheral system components, especially once SOEC stack costs have declined. Therefore, vertical scaling in larger installations becomes a key lever for SOEC system CAPEX reduction by enabling centralized, more efficient infrastructure and better utilization of fixed assets.[7, 8]

A recent study estimated a CAPEX of 950 $/kW to be possible for a GW SOEC system with today's technology assumed to be manufactured at the GW scale.[8] In such a system the BoP components can represent 85 % of the overall system cost, partly due to the high cost of low-volume BoP components. As a result, identifying and leveraging cross-industry synergies, for

instance with applications in concentrated solar power and industrial heat, could be key to reduce BoP component costs through standardization.

At the plant level, previous public estimates have frequently underestimated hydrogen production plant CAPEX by omitting project and location related costs such as buildings, electrical infrastructure, indirects, and contingencies.[211] As a result, entire electrolyzer systems were suggested to rather represent a type 2 technology, that is more design-intensive and customized leading to lower learning rates via indirect learning.[211] Effective cost reductions for Type 2 electrolyzer systems can be achieved by leveraging cross-industry synergies, standardizing components, and integrating balance-of-plant designs, enabling system costs to decrease alongside stack learning curves.

### *7.3 Operational expenditure (OPEX) and LCOH*

The ongoing economic performance of an SOEC plant is determined by OPEX. OPEX is typically dominated by electricity costs, but also encompasses other contributions such as routine maintenance, reactant consumption and stack replacements. Together with annualized CAPEX converted into an equivalent yearly cost using the capital recovery factor (CRF), which accounts for project lifetime and cost of capital, these cost contributions define the LCOH as follows

$$\text{LCOH} = \frac{CAPEX*CRF+(Electricity\ Price+Grid\ fees)*Annual\ Electricity\ Consumption\ +Other\ OPEX}{Annual\ Hydrogen\ Production}. \quad (10)$$

For SOECs, electricity contributes 1.2-3.8 €/kg$_{H_2}$ at electricity prices of 30–100 €/MWh, assuming a specific energy consumption of 38 kWh/kg$_{H_2}$. This is notably lower than the 1.5-5 €/kg $H_2$ estimated for low-temperature electrolysis technologies (~50 kWh/kg$_{H_2}$). Besides, grid fees of typically 5–25 €/MWh are added for grid-connected electrolysers but can be lower or negligible for projects co-located with dedicated renewables. The lower energy intensity of SOEC systems provides a clear economic advantage in electricity-dominated cost structures.[214] Economically, hydrogen plants should be put on standby when high electricity prices lead to a strong increase in product price. Since SOECs have a higher break-even than AEL or PEMEL, they can operate longer under variable electricity pricing, increasing operational hours and potential earnings.[187]

Another OPEX contribution is the provision for stack replacements, reflecting uncertainties around long-term durability. Current stack lifetimes are roughly five years,[105] and hence, stacks must be replaced 4-5 times over a 20-25 year plant lifetime. Compared to other electrolyzers, this issue is particularly pronounced since SOEC have still a limited track record with few >10 kh degradation studies available, especially in the field.

High electricity costs and utilization rates above 4,000 full-load hours per year substantially diminish the contribution of CAPEX to LCOH. [214, 215] Ceres has reported projected LCOH of ~4.90 €/kg with FOAK, with a potential decrease to ~4.30 €/kg in a hypothetical ~60 MW SOEC system operated at electricity costs of ~68 €/MWh.[105] Even with a system CAPEX of 2000-2500 €/kW, an LCOH of 7-8 €/kg was reported to be feasible.[216] Some projections indicate that LCOH will fall to 2-3 €/kg by 2030.[208] However, recently an increasing number of reports suggested that these numbers could be overly optimistic.[211, 214, 217] Boston Consulting Group predicted that, irrespective of the electrolyzer technology used, LCOH in Central Europe will more likely be in the range of 5-8 €/kg by 2030 due to a deteriorated macroeconomic context, higher energy market prices and supply chain challenges for both wind power and electrolyzers.[217] Another recent report based on real electrolysis projects in the Netherlands showed that plant CAPEX today is significantly higher than generally assumed partly due to indirect costs leading to LCOH of 6-13 €/kg.[218]

Currently, the cost of grey hydrogen is ~3 €/kg,[217] and as a result, there remains a large gap to be bridged for green hydrogen to be competitive. The gap could be reduced by emerging policies, such as the potential RED III penalties in Europe, but LCOH in Europe would need to be below 5 €/kg to avoid being threatened by imports.[217]

# 8 Scale-up strategies and modularization

Modularization enables plant scale-up to large capacities via horizonal scaling by numbering up of self-contained SOEC systems. This approach offers several advantages compared to conventional centralized plants, such as reduced financial and operational risk through distributed production, increased flexibility to changing demand, short construction times through factory manufacturing, low transportation and upfront engineering costs.[219] However, the unit sizes of state-of-the-art cells, stacks and stack modules are currently still relatively small compared to the low-temperature electrolysis technologies which has economical drawbacks. Significant cost reductions can be achieved via vertical scaling, that is, by enabling higher current densities,[208] and increasing the sizes of cells, stacks and stack modules. Additional reductions can be realized by expanding manufacturing capacity through enhanced automation.[7, 8]

State-of-the-art stack technologies currently mostly reach current densities of ~0.5-0.6 A $cm^{-2}$ in ESCs, ~0.6-0.8 A $cm^{-2}$ in FESCs, and ~0.25-0.4 A $cm^{-2}$ in MSCs at nominal, isothermal operation.[38, 57, 186] An increase of operating current density is targeted by most manufacturers, regardless of the stack architecture. For example, Sunfire has developed a new ESC-based stack technology which can increase the current density to 0.8-1.1 A $cm^{-2}$.[185]

Most stack designs currently employ cells with an active area of ~100 $cm^2$ (Table 1). These designs were frequently initially developed for SOFC operation where temperature gradients are high and large cells and stacks increase the risk of mechanical failure. Furthermore, large ceramic cells are also difficult to manufacture with a high production yield. However, temperature gradients are generally lower in SOEC operation, particularly at isothermal operation, and progress in cell and stack manufacturing processes has been made. As a consequence, cells with active areas of 200 $cm^2$ up to 900 $cm^2$ have been manufactured and successfully tested in stacks.[220, 221] In their TSP-2 stack platform, Topsoe is using single cells with a presumed active area of 300-400 $cm^2$ and a hexagonal shape.[58] Ceres Power has also recently increased the active area of their cells from 80 $cm^2$ to 240 $cm^2$.[39] An alternative to manufacturing cells with a large area is the window-frame design, for example used by SolydEra, which enables the integration of a larger number of smaller cells in one plane minimizing the risk of cell fracture and reducing steel and sealing material per kW.[52, 222]

Stacking a large number of cells requires stringent manufacturing tolerances in order to maintain operational and structural integrity. Large component variations can accumulate and result in contacting issues between cells and interconnects, or lead to internal or external leakages.[223] A mechanical load is frequently applied on the stacks to avoid such issues, for instance, via compression springs. Some designs include buffer plates for additional stress managament.[63] While earlier stack designs were limited to around 30 cells,[63] newer concepts frequently contain more than 100 cells in a stack (Table 1).

Larger stack modules have the advantage of centralized BoP components if cost-effective products are available in the respective sizes. With regards to the degree of centralization, different approaches can be followed. For instance, Topsoe's 350 kW electrolysis hotbox includes a heat exchanger and electric heater,[186] whereas Ceres is currently working on a hotbox with up to 725 kW power consumption in a pressure vessel of up to 10 bar that excludes all heaters and HEXs to achieve a higher degree of centralization.[154] Larger hotbox size can also decrease the required amount of insulation material due to lower specific surface areas. Both measures can lead to cost reductions. However, large hotboxes face an increased risk of inhomogeneous fuel and air distribution. Moreover, a larger number of electrically connected stacks increases the undesired downtimes in case of the damage in a single cell or stack. In this regard, efficient replacement strategies for single stacks are required.

So far, stacks are typically connected in series within a stack module which increases the total voltage at constant current. However, according to International Electrotechnical Commission (IEC) standards, the low-voltage limit in Europe is 1500 $V_{DC}$ or 1000 $V_{AC}$, and higher voltages impose stricter safety requirements. Electrical feedthroughs at high voltage and high temperatures pose a safety risk due to possible gas leakages and creepage currents that require materials with a high dielectric strength. As a result, most stack module manufacturer have so far remained below this limit. Given the substantial cost contribution of power electronics, manufacturers aim to reduce cost by leveraging the existing economies of scale from the more mature photovoltaics (PV) or wind turbine industries. Both SOEC AC/DC converters and wind/solar DC/AC inverters use similar power semiconductor technologies and control electronics, including such as DC/DC converters, enabling shared components and design platforms. In this regard, module sizes of 1-5 MW seem to be attractive to minimize power electronics costs.[211]

An alternative reported strategy is to operate multiple stacks in parallel at a lower total voltage with separate power electronics. Such an electrical interconnection of stacks in a module can be used to stay at lower voltages and achieve power electronics operation at highest efficiencies.[102]

System sizes in the MW range are already available, for instance 1.2 MW by Bloom Energy.[104] Convion is currently working on the development of a versatile 1 MW system that can also be operated reversibly based on stacks from SolydEra or Elcogen.[150] FuelCell Energy is developing a 1.1 MW system prototype.[47] In 2024, Sunfire has started to advertise their SOEC system with 10.7 MW power.[180] Moreover, Ceres and Shell have started to develop a pressurized 10 MW system design.[224]

SOEC systems are generally assumed to have a large physical footprint. However, technological progress is evolving rapidly and Sunfire specifies a footprint of 24 $m^2$/MW for their new 10.7 MW HyLink system,[180] which is significantly lower than the reported 300 $m^2$ for their previous 2.6 MW installation in Rotterdam. The new Sunfire SOEC system even has a lower specific footprint than the required 41.5 $m^2$/MW for their 10 MW AEL system.[103]

## 9 Installations and manufacturing capacities

SOECs are currently predominantly installed with capacities between 100-500 kW, and only recently installations in the multi-MW scale have been deployed.[225] Bloom Energy has put a 4 MW solid oxide electrolyzer into operation in 2023 at NASA's Ames Research Center in Mountain View, California, USA.[226] In the European MultiPLHY project, Sunfire has installed a 2.6 MW electrolyzer plant in the Neste biorefinery in Rotterdam, Netherlands in 2023,[227] and within the GrinHy 2.0 project a 720 $kW_{AC}$ electrolyzer in Salzgitter, Germany in 2022.[228] Shell and Ceres have deployed a 1 MW SOEC demonstrator in Bangalore, India.[229] Moreover, Ceres is working together with Linde Engineering and Bosch on the installation of a 1 MW demonstration plant in Stuttgart, Germany.[230] Dynelectro is planning to operate a 1 MW SOEC system with stacks from SolydEra in 2025 with their patented AC:DC operating technique.[231] The tubular design is developed by Mitsubishi Heavy Industries and a 400 kW pressurized system was taken in operation in Takasago City, Japan in 2024.[232] Based on these demonstration projects, it can be concluded that SOECs for hydrogen production has currently reached TRL 8. Topsoe's SOEC technology is being deployed at an approximately 100 MW ammonia plant at the Port of Victoria in Texas. This marks the first deployment of SOEC technology beyond the 10 MW scale, with commercial operations targeted for 2027.[233]

In the German Kopernikus project P2X, a co-electrolysis stack module was deployed and operated in 2022 with an achieved power of up to 220 kW to produce syngas and produce synthetic fuels downstream putting the technology at TRL 5-6.[234] Topsoe has leased two eCOs units with each ~340 kW for $CO_2$ electrolysis to DeLille Oxygen Co. (USA) already in 2020 elevating the technology to TRL 8.[235] Despite these advances, long-term performance and degradation data of these systems are limited, and such results will play a key role in de-risking investments and accelerating commercial deployment.

So far, SOEC production capacity is still performed mainly at pilot line scale, although. multiple large production factories have been announced. Bloom Energy already approaches SOEC manufacturing capacity of over 2 GW/year.[104] Topsoe inaugurated a 500 MW/year plant in Herning, Denmark in 2025.[236] Elcogen opened a production line with an SOEC capacity of 360 MW/year in Tallinn, Estonia, also in 2025.[237] Thyssenkrupp nucera is building a 300 MW/year SOEC production plant based on the licensed technology by Fraunhofer IKTS in Germany.[238] SolydEra currently operates a 75 MW/year industrial stack production plant in Pergine Valsugana, Italy.

## 10 System integration with exothermic downstream processes

The global hydrogen market is evolving rapidly, yet the future role of green hydrogen remains uncertain, and depends heavily on the costs for its production, storage and distribution.[239] Despite these uncertainties, several frameworks identified high priority end-use applications for hydrogen,[3, 240] including fertilizer production (ammonia), shipping (ammonia, methanol, methane), aviation (FT-derived kerosene), and the steel and chemical industries (methanol, methane).[241] The underlying production processes involve highly exothermic processes, offering strong synergies for SOEC integration, by enabling the use of excess heat for steam generation,[143] significantly improving overall system efficiency. This chapter explores these synergies, discussing the benefits, challenges, and technical considerations of integration.

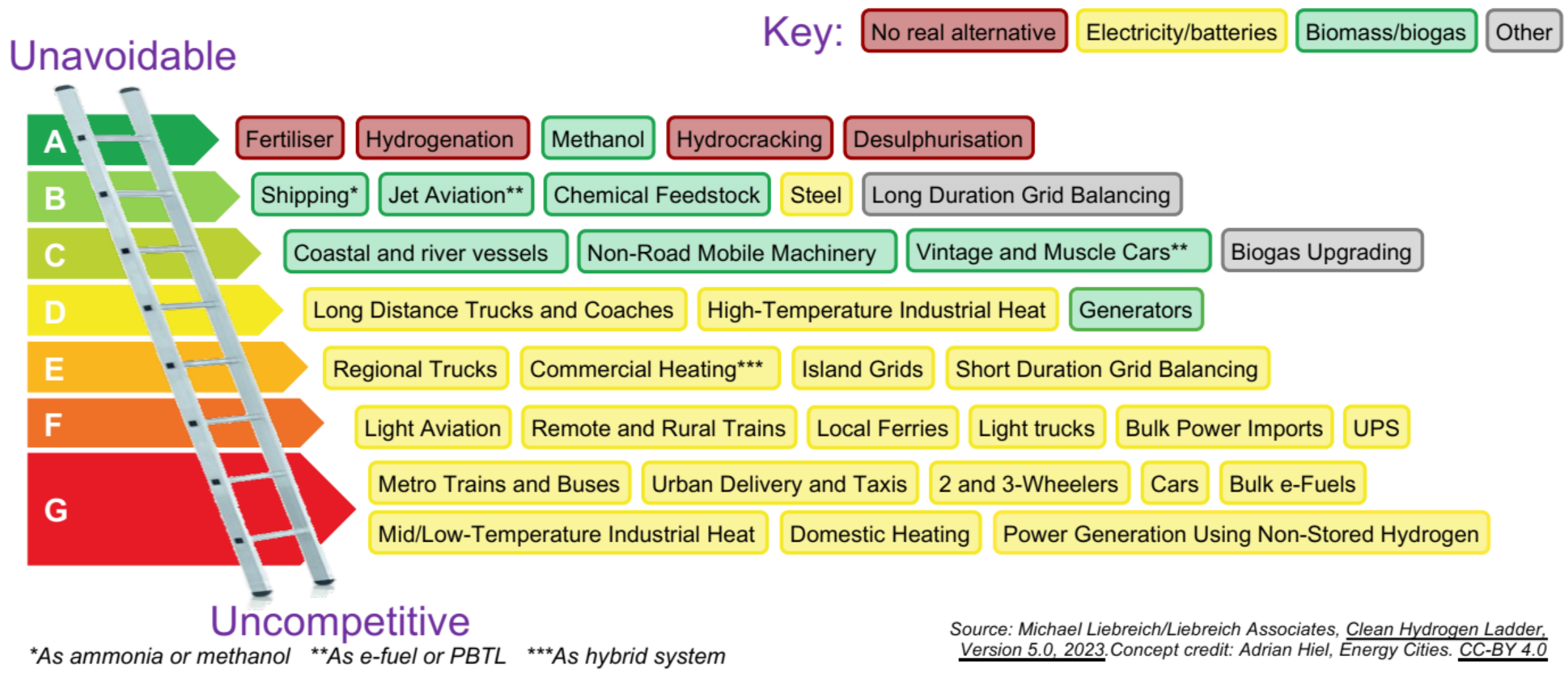

Figure 13. Hydrogen Ladder 5.0, an illustration showing the likelihood for different use cases being significant hydrogen users considering costs, safety, convenience, critical mineral availability, co-benefits, externalities, geopolitics, thermodynamics, physics, chemistry, and economics. It does not include efficiency and market size. Source: Michael Liebreich/Liebreich Associates, Clean Hydrogen Ladder, Version 5.0, 2023. Concept credit: Adrian Hiel, Energy Cities. CC-BY 4.0.[240]

### *10.1 P2X with chemical synthesis processes*

Key operating parameters for state-of-the-art methane, methanol, ammonia and FT synthesis reactors are summarized in Table 3. Thermodynamic calculations indicate that the combined heat from the exothermic synthesis reactions and compression of SOEC product gas could theoretically supply the entire 40.6 kJ/mol required for steam generation at 1 bar in most cases (Figure 14). However, heat exchanger losses and additional heat demands from auxiliary process units reduce recoverable heat in practice. Therefore, effective thermal integration of SOEC, downstream reactor and auxiliary units requires exploration via pinch analyses using detailed process & instrumentation diagrams (P&ID).

All discussed reactions are exothermic, with equilibrium conversion favored by low temperatures and high pressures due to a net reduction in gas molecules. Furthermore, heat removal and gas recycling are crucial for high conversion due to the exothermicity of the reactions. Typically, a purge gas stream is essential to prevent the accumulation of inert gases and impurities, which can increase reactor size, negatively affect reaction kinetics and degrade catalyst performance. High-purity hydrogen produced by SOEC could reduce the purge gas stream.

Most industrial reactors operate under steady-state conditions and are optimized for yield, selectivity, and catalyst longevity. Multi-bed, adiabatic fixed-bed reactors with intercooling and gas recycling are most common for all synthesis reactions due to their simplicity.[242] More advanced reactor types, such as micro-structured,[243, 244] fluidized-bed, or slurry-bed reactors, [244-246] are sometimes used for improved heat management and near-isothermal conditions. Due to the expected future increase in RES share, dynamic operating of synthesis reactors and the entire P2X process chain is gaining importance. Flexible, modular reactors with improved load-following capabilities are being developed, particularly for decentralized or off-grid applications.[247]

The following subsections summarize each synthesis process and their integration potential with SOECs, focusing on operating regimes, mass and energy flows, heat integration and dynamic behavior. More comprehensive information of the different synthesis processes can be found elsewhere.[242, 248]

Table 3. The most promising industrial downstream synthesis reactors for SOEC coupling and their reaction enthalpy, temperature, pressure and favored catalyst.

| Reaction | ΔH°r / kJ·mol–1 | Temperature / °C | Pressure / bar | Catalyst | Single pass conversion | Inlet gas composition |
|---|---|---|---|---|---|---|
| Ammonia synthesis: $3H_2 + N_2 \rightarrow 2NH_3$ | -105.9 (450°C) | 350-500 | 100-300 | Fe-based | 20-35%[249, 250] | $H_2$/N2 = 3, $H_2O$ <1-5 ppm |
| Methanation: $3H_2 + CO \rightarrow H_2O + CH_4$ $4H_2 + CO_2 \rightarrow H_2O + CH_4$ | -218.2, -178.6 (350°C) | 200-550 | 4-100 | Ni-based | 60-99 % [243] | $\frac{H_2 - CO_2}{CO + CO_2} \sim 3$ |
| FT synthesis: $(2n+1)\ H_2 + n\ CO \rightarrow C_nH_{2n+2} + n\ H_2O$ | -162.2 (250°C, n=10) | 180-260 | 15-60 | Fe-based | 60-85%[242] | H2/CO = 2 |
| Methanol synthesis: $2H_2 + CO \rightarrow CH_3OH$ $3H_2 + CO_2 \rightarrow CH_3OH + H_2O$ | -98.3, -58.0 (250°C) | 200-300 | 50-100 | $CuO/ZnO/Al_2O_3$ | CO-rich: <80 %, $CO_2$-rich: <45 %[251] | $\frac{H_2 - CO_2}{CO + CO_2} \sim 2$ |

Table 4. Main flexibility parameters for the large-scale downstream synthesis reactors.

| Reaction | Load range | Ramp rate | Hot start-up | Cold start-up |
|---|---|---|---|---|
| SOEC system | 0-100 %[47] | 10 %/min[47, 104, 180] | 10 min[104] | ~24 h[252] |
| Ammonia synthesis | Traditional design: 65-100 %;[253, 254] | ± 20%/h[258] | - | Several days[250] |
| | Modified designs: 10-100 %[254-257] | | | |
| Methanation | Adiabatic fixed bed reactors: 40-100 %,[259] Three-phase and cooled fixed-bed reactors: 10-100 %[243] | 3-4 %/min [259, 260] | - | - |
| FT synthesis | - | - | - | - |
| Methanol synthesis | 15-100 %[242] | 5-13 %/min [242] | - | - |

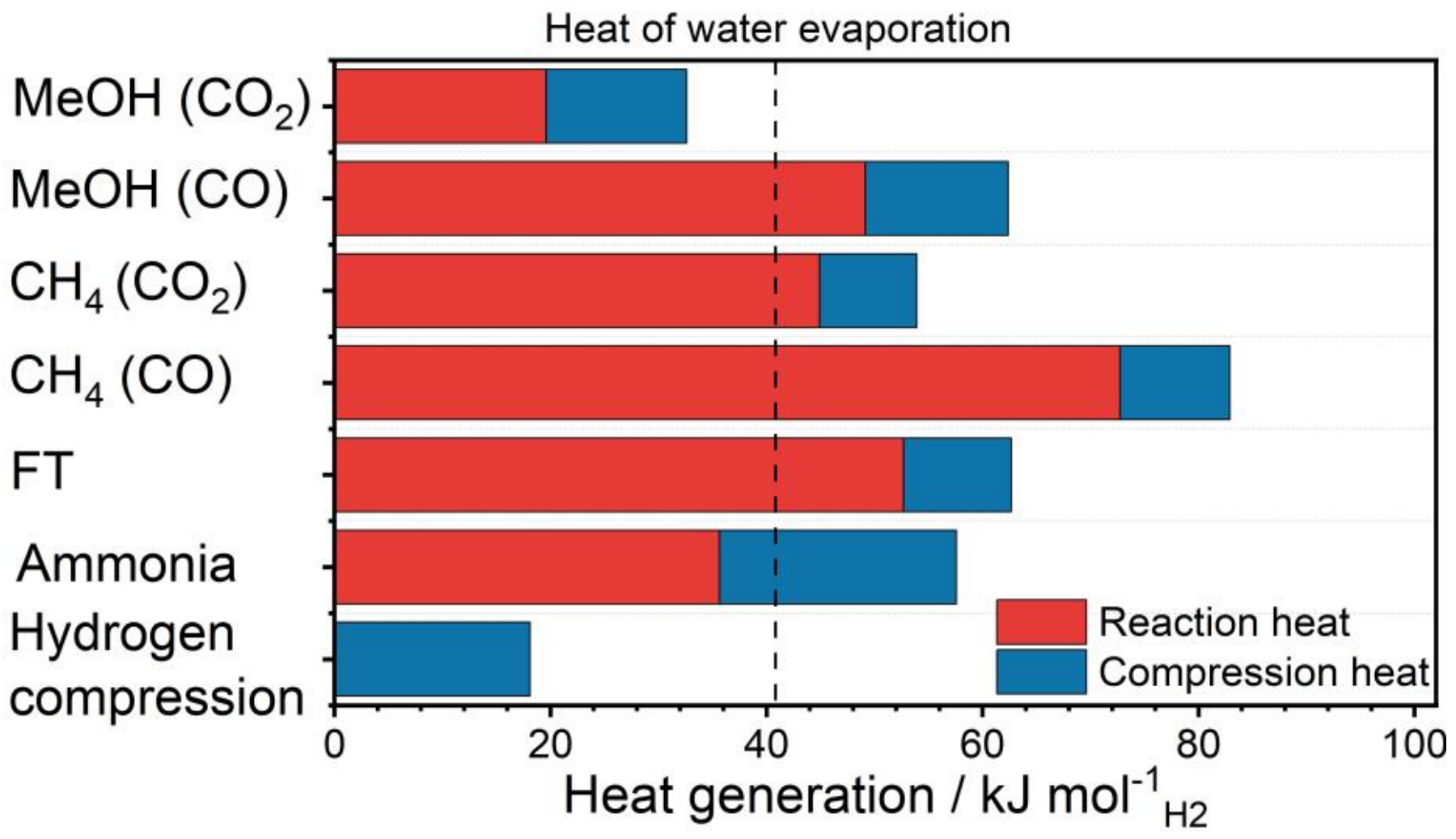

Figure 14. Thermodynamic assessment of heat generation per mole $H_2$ during reaction and compression for different downstream processes. Compression heat generation was calculated based on the assumption of isentropic compression with a compressor efficiency of 100% and maximum gas temperatures of 200°C.[106] Considered processes were methanol synthesis (75 bar, 250°C, 68.5% $H_2$, 26.5% CO, 5% $CO_2$), methanation (25 bar, 350°C, 75% $H_2$, 25% CO), FT synthesis (30 bar, 250°C, 67% $H_2$, 33% CO), ammonia synthesis (200 bar, 450°C, 75% $H_2$, 25% $N_2$) and hydrogen compression (300 bar, 100% $H_2$). A reactant conversion of 100% was assumed in the reactors. The dashed vertical line indicates the heat of evaporation of water.

#### *10.1.1 Power-to-Ammonia*

Ammonia is the second most produced chemical globally, with an annual capacity of ~180 Mt, consuming nearly 2% of global energy. Primarily used in fertilizers, it also serves as a chemical feedstock and a potential maritime fuel. The conventional Haber-Bosch (HB) process has been optimized over a century and integrates a cryogenic air separation unit (ASU), methane steam reforming for hydrogen, and the HB synthesis loop.

*Integration with SOEC*

Replacing steam reforming with SOEC in existing (brownfield) HB plants is attractive but complex due to already existing high levels of system integration. Commercial adiabatic HB reactors use multiple catalyst beds and intercooling,[250] achieving 20–35% single-pass conversion. Ammonia is separated by condensation, and unconverted gas is recycled for a high yield.[249, 250]

Currently used large-scale ammonia plants are designed for steady-state operation with a nominal capacity of 1000-3000 t/day.[253, 261] The smallest practical size of an ammonia plant with the conventional design is 250-300 t/day, which might be better suited for dynamic operation.[253, 261] The replacement of the steam reforming unit in such a case would still require an SOEC system with an estimated power of >100 MW.[261]

The ammonia synthesis reaction provides up to ~31 $kJ \cdot mol_{H2}^{-1}$ heat, insufficient to cover the entire heat demand of steam generation (Figure 14). For isothermal steam electrolysis, at 1 bar, process simulations showed 57% of steam demand can be met using HB reactor and intercooler heat, versus only 31% at 30 bar due to the unavailability of sufficient high-temperature heat at the elevated water evaporation temperatures of 234°C.[145] Additional steam can be generated using SOEC off-gas heat, but electric heaters would still be required to provide the remaining heat of evaporation. Alternatively, exothermal SOEC operation could close the steam balance.[262] Reported overall system efficiencies of 71-74 % exceed those of AEL-based systems by ~17 %.[145, 262]

*Flexibility and Dynamic Operation*

The established HB process with multiple high-pressure compressors and complex heat integration is difficult to operate dynamically.[263] Large-scale plants allow operation at 65-100% of the nominal capacity due to minimum flow rate requirements to avoid overheating.[253, 254] Some companies have proposed approaches to decrease the minimum load, for instance by adjusting recycling rates[264, 265] or adapting the HB heat exchanger design for improved heat integration at part load.[255, 256] Reported minimum operating limits are as low as 10%, [254, 257] however, more realistic values are arguably significantly higher.[266] Ramp rates of ± 20%/h were reported.[258] Cold start-up of large-scale reactor can take several days due to the complex start-up procedure for purification of the reactants as well as the cryogenic ASU due to the required reduction of the Fe-based catalysts.[250]

At smaller HB reactor capacities of ~250 t/day, heat losses and energy consumption increase significantly.[249] Significant research is focused on the development of medium-pressure (~20 bar) ammonia synthesis reactors using highly active Ru-based catalysts for flexible, small-scale applications.[267, 268] These systems benefit from improved ramping capabilities and reduced energy consumption when combined with absorption-based ammonia separation.[263] However, regeneration of the absorption material may add heat demand,[269] and further material innovation is still required.[249]

*Alternative system concepts*

Cryogenic ASU is an energy-efficient technology to produce nitrogen at large scales (>100 MW), but generally does not scale down well.[270] By contrast, PSA plants are less energy-efficient, but more scalable, allow start-up and shutdown within minutes and can provide a flexible flow rate following demand.[269] Thus, the combination of PSA, SOEC and small, decentralized HB reactors could provide a more flexible alternative for dynamic operation.

Topsøe has patented a concept using SOECs as oxygen permeation membrane and eliminating the CAPEX related to the installation of either ASU or PSA.[271, 272] According to this approach, a cascade of SOECs in endothermal operation is employed and air is introduced into the feed gas at several points for partial hydrogen combustion, simultaneously introducing nitrogen into the feed gas and providing heat for feed gas preheating. No electric heater is required during operation to close the steam balance. A demonstration plant with 500-1000 kg/d is planned in 2025.[235] First Ammonia (USA) is building a ~100 MW ammonia plant in Port of Victoria (Texas, USA) using Topsoe's SOEC technology. Commercial operation is targeted for 2027.[233]

### *10.1.2 Power-to-Methane (PtM)*

Power-to-methane (PtM) systems convert CO/$CO_2$ and $H_2$ into methane through electrolysis and catalytic methanation, e.g. for long-term energy storage.[273] Steam electrolysis could be combined with a $CO_2$ methanation reactor, but the production of CO-rich syngas via co-electrolysis and coupling with a CO methanation reactor offers several benefits.

Low temperatures promote the conversion of both CO and $CO_2$ (Figs. 7&8) and thus, are desirable to obtain high-purity methane. However, temperatures of 200-550°C and the Ni-based catalysts are required to achieve high reaction rates. Careful temperature control is key to prevent sintering of the Ni catalyst, carbon formation via methane decomposition, and achieve the desired product gas composition. Relatively high single pass conversion values of >85% can be achieved due to fewer competing side reactions and the high exothermicity. For CO methanation, conversion can exceed 95% without recycling, significantly reducing CAPEX.[244] Pressures of 5-100 bar are used to achieve high methane formation rates,[243] although this requires significant energy for compressing the inlet gas. The feed gases of the methanation reactor should have $H_2$/CO > 3 and $H_2/CO_2$ >4 to ensure an over-stoichiometric hydrogen concentration.[274] Higher hydrogen concentrations diminish the probability of carbon formation and lead to higher $CO/CO_2$ conversion, but result in hydrogen remaining in the product gas.[275] Methane membrane separation is technically challenging, but residual hydrogen in synthetic methane can be tolerated in certain applications, such as Germany's natural gas grid with up to 10 vol % of hydrogen. Some residual steam in the SOEC off-gas may be helpful to reduce the risk of coking.

*Integration with SOEC*

CO methanation provides up to 71.9 $kJ \cdot mol_{H2}^{-1}$ heat at 350°C, more than sufficient for SOEC feed water evaporation (Figure 14). Indeed, process simulations including heat integration indicate that no electric heaters are required for steam generation in a PtM system based on isothermal co-electrolysis.[276] However, more heat is required if hydrogen is produced via steam electrolysis and $CO_2$ is only added in the methanation reactor (Figure 14). In such a case, system efficiencies decrease due to electrical energy consumption of the heater.[277] Exothermic operation could help to cover the additional heat demand in such cases. Pressurized co-electrolysis could increase system efficiency by enabling methane formation already within the SOEC, reducing specific energy demand for compressors, SOEC, heaters and evaporators due to decreased specific water inlet flow.[106] Reported optimized integrated PtM efficiencies range between 80-85 %.[163, 276, 278] This is significantly higher than the largest PtM installation to date, the 6 MW Audi e-gas plant in Werlte, Germany, that uses PEM electrolysis, but achieves only 54 % system efficiency.[279]

*Flexibility and Dynamic Operation*

Transient operation can induce significant temperature changes in the methanation reactor. For example, a change from the full design load of a catalytic fixed bed reactor to the 20 % part load point can cause a temperature decrease by more than 100 K and a simultaneous hydrogen conversion decrease from 99.9 % to 98.0 %.[280] Reactor recycling can help buffer these temperature changes.[260] Ramp rates of 3-4 % $min^{-1}$ were reported to be possible in fixed-bed reactors without recycle without having detrimental effects on gas quality.[259, 260] Fixed-bed reactors typically require a minimum load of 40%,[259] although three-phase reactors can operate as low as 10%.[243] During start-up, standby and shut-down of the methanation reactor reducing conditions are required to avoid Ni oxidation,[281] and synergies with reducing gas required for Ni/cermet fuel electrode operation could be exploited.

*Demonstration Projects and Industrialization*

A power-to-methane system based on SOEC was implemented in the European HELMETH project.[280] A 10 kW steam electrolysis stack module pressurized at 8 bar from Sunfire was integrated with a fixed-bed $CO_2$ methanation reactor. Heat integration was achieved by using waste heat from the methanation reactor to directly generate steam at 250°C and 40 bar which was subsequently expanded to match the SOEC operating pressure.[280, 282] Furthermore, Topsoe carried out a demonstration project for biogas upgrading to biomethane, using a steam electrolyzer and a full-scale methanation unit.[283, 284] Based on a similar process, Reverion GmbH (Germany) is commercializing a 500 kW reversible SOC system for biogas uprading and electricity generation. However, no direct coupling of co-SOEC and CO methanation units has been done on an industrially relevant scale yet.

*10.1.3 Fischer-Tropsch synthesis*

FT synthesis converts syngas into synthetic hydrocarbons and has been commercially established since the 1950s (Table 3). Its product spectrum contains alkanes, alkenes and oxygenates, with chain length distribution determined by the probability of chain growth α. Low α values favor gaseous products ($C_1$–$C_4$), while higher values yield liquid fuels ($C_5$–$C_{19}$) and heavy waxes ($C_{20}$+). Product distribution depends on catalyst, temperature, pressure (typically 15–60 bar), and feed gas composition.

High-temperature FT (300–350°C, Fe catalysts) favors gaseous products, while low-temperature FT (180–260°C, Co catalysts) produces longer-chain liquids and waxes, preferred for power-to-liquid (PtL) applications since liquid products with C10-C23 can be used as substitute gasoline or diesel. Wax products may require hydrocracking to upgrade to naphtha, diesel and jet fuel. Higher temperatures lead to the frequently undesired formation of short-chain hydrocarbons and methane, and possibly to accelerated catalyst degradation and carbon deposition. Low-temperature FT requires a CO-rich syngas mixture, whereas the process based on $CO_2$-rich syngas gas is still in early development. As a result, coupling pure water or steam electrolysis with low-temperature FT necessitates a dedicated RWGS reactor, while coupling with co-electrolysis can provide unique benefits by simplifying process design. A $H_2$/CO ratio of ~2 is ideal. Industrial FT reactors operate at 60-85 % single pass conversion to prevent hot spots, and different recycling approaches can be considered (Figure 15).[242] Internal FT recycling increases syngas conversion, but lowers the yield of higher hydrocarbons and a high purge gas stream to avoid accumulation of inert gases and undesirable gaseous C1-C6 by-products.[285]

*Integration with SOEC*

External 'long' recycle of the FT tail-gas into the SOEC could improve process and carbon efficiency, [147, 286, 287] due to the reforming of gaseous hydrocarbons and reduction of $CO_2$ on Ni/cermet fuel electrodes and their recycling into the FT reactor.[247] It has been demonstrated on lab scale that in such a case no short recycling of SOEC product would be required.[286]

However, the endothermic reforming process could induce undesired thermal gradients in the stack and the risk of coking. A separate steam reforming reactor could be included in the FT recycle stream to reform hydrocarbons and avoid strong cooling and cooking. More work is required to investigate the influence of hydrocarbons on co-electrolysis performance and durability to assess the feasibility of long FT tail-gas recycling.

The FT reaction heat can be used for SOEC steam generation,[147, 288] however, a significant portion is already consumed by the energy-intensive product distillation in existing plants. Additional heat for raising the recycle stream to SOEC operating temperatures could be supplied by purge gas combustion.[288] Depending on the used system layout, concept studies have reported integrated co-electrolysis-based liquid fuel production efficiencies between 51-63 %.[147, 288, 289]

Residual ~10 % $CO_2$ in the dried co-SOEC product gas may lead to increased compressor energy consumption, lower reaction rates in the FT reactor and increased costs for piping. The removal of $CO_2$ via TSA may be a viable alternative process route and has not been explored in detail yet.

*Flexibility and Dynamic Operation*

Operating FT synthesis reactors dynamically is challenging due to the complex and not fully understood reaction mechanism, making reliable kinetic modeling difficult.[290] Little has been published with regards to concrete values for load range and minimum load. However, dynamic operation as a tool for process intensification has experimentally investigated in the past.[290] Especially periodic hydrogen exposure was reported to help drain the catalyst bed and remove accumulated liquid products from the catalyst pores with the result of restoration of initial catalyst activity.[291, 292] However, this can lead to an increased short-chain hydrocarbon selectivity.[292, 293] This suggests the potential to use hot-standby times of the SOEC for FT synthesis catalyst regeneration via continued recirculation of reducing fuel gas. Operation of FT synthesis at constant temperature will lead to variations in conversion rate and selectivity. In micro-structured reactors, operation at constant conversion with reduced methane formation was achieved by controlling the temperature via pressure-regulated controlling.[294, 295] Similar to SOEC fuel electrodes, FT reactor start-up and shutdown are typically carried out with reducing gas to maintain the Co catalyst in a reduced state allowing the continuous fuel gas circulation of reducing gas through both reactors during standby.

*Demonstration Projects*

In the Finnish e-Fuel project in 2024, a 130 kW steam electrolyzer by Convion was coupled to a mobile FT unit including a RWGS reactor in 2024, producing 300 kg FT crude product.[296] In the German Kopernikus P2X project, a 10 kW co-electrolyzer from Sunfire, microstructured FT and a hydrocracking unit were integrated in 2019.[102] In 2024, the follow-up project P2XII scaled up to a 250 kW co-SOEC-based PtL plant. Based on Sunfire's solid oxide and alkaline electrolysis technologies, Norsk e-fuel plans large-scale e-fuel production using SOEC in Mosjøen, Norway with a production volume of 25 million t/year.[297]

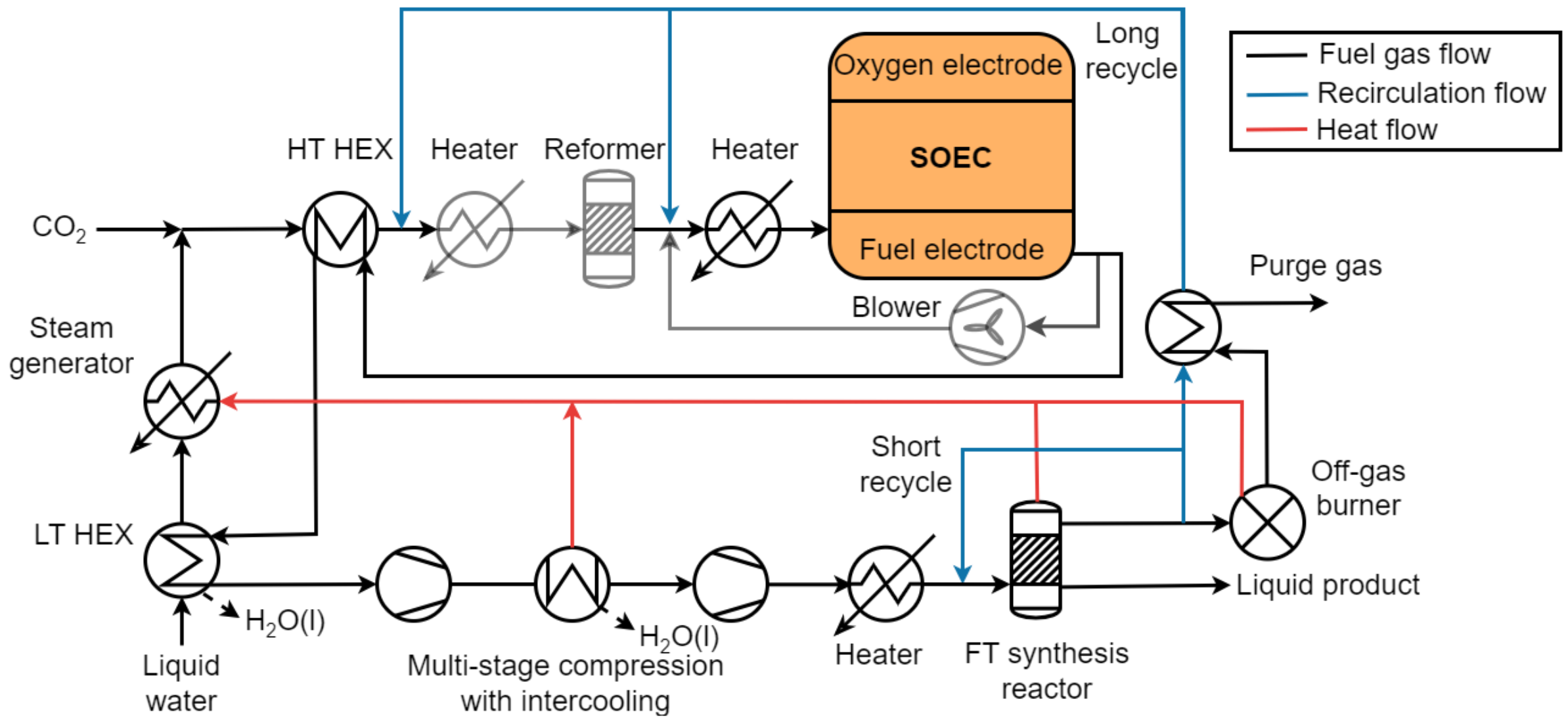

Figure 15. A simplified PFD depicting the coupling of a co-electrolysis module with a FT synthesis unit featuring a long and a short off-gas recycle (depicted in light blue). A similar PFD can also be applied for co-SOEC integration with a methanol synthesis reactor. Gas flows on the oxygen side of the SOEC are neglected for enhanced visibility. Electrical heater and reformer before the SOEC are depicted in low opacity to indicate their optionality in a long recycle. HT blower is shown in low opacity since it may not be required in a long recycle.

#### *10.1.4 Power-to-Methanol*

In the industrial methanol synthesis process, a $H_2/CO/CO_2$ feed gas derived from natural gas is converted into methanol at 50-100 bar and 200-300°C. Using $CuO/ZnO/Al_2O_3$ catalysts, selectivity exceeds 99% under ideal conditions.[242] However, sub-optimal conditions can lead to higher alcohol formation. The process is thermodynamically limited by high $CO_2$ concentration, restricting single-pass conversions to below 45%.[242, 251] In contrast, the $H_2/CO$ synthesis reaction has less limitation, allowing conversions of up to 80 % (Table 3). However, $CO_2$ concentrations of 2-4 % are kinetically favorable since at lower $CO_2$ partial pressures, the oxide catalysts are over-reduced and at higher concentrations the catalyst surface is deactivated.[298, 299] The feed gas composition is defined by the stoichiometric number (SN),

$$SN = \frac{H_2 - CO_2}{CO + CO_2},$$

where the molar fractions of the species are used. The value of SN is ideally slightly above 2 for maximum conversion.[242] The kinetics of methanol synthesis are slower than methanation and FT synthesis, allowing quasi-isothermal operation in simple fixed-bed reactors despite the process's exothermic nature.[300]

Most power-to-methanol development has so far focused on $CO_2$-based feeds with hydrogen produced by low-temperature electrolysis.[242] However, this can lead to crude methanol with up to 20 % more water than in reactors with CO-rich syngas.[242] Furthermore, single-pass conversion is low due to thermodynamic limitations, requiring high recirculation rates and larger reactors. Excess water in the reactor can also reduce catalyst activity and lifetime due to steam-induced aging.[301, 302] Retrofitting existing methanol plants to the direct $CO_2$ pathway requires adjustments at multiple stages.

*Integration with SOEC*

In contrast, co-electrolysis allows for simpler integration with methanol synthesis reactors at higher yields. Unlike $CO_2$ methanolization, CO methanolization can theoretically supply sufficient heat for SOEC feed water evaporation (Figure 14). Moreover, final product purification through distillation is energy-intensive, consuming much of the generated reaction heat, limiting excess heat available for SOEC integration. For steam electrolysis coupled with $CO_2$ methanolization, the waste heat covers only 42 % of the heat required for water evaporation.[303] In co-electrolysis, excess heat can fully meet the steam generation demand if the SOEC is operated exothermally,[304] improving the integrated process efficiency.[305] Reported power-to-methanol system efficiencies range from 70-80%.[305, 306]

Methane is inert in the methanol synthesis and its accumulation in the reactor is a concern, especially at high SOEC pressures where methanation is favored. Hence, co-electrolysis-based process efficiency is highest at atmospheric SOEC pressures.[304] A long recycling approach, as proposed for FT synthesis in Figure 15, could mitigate methane buildup in methanol reactors, though this system concept has not yet been explored.

Industrial methanol synthesis reactors can be ramped faster than other downstream reactors due to efficient temperature control (Table 4). However, the distillation unit is less adaptable, requiring crude methanol buffering and advanced process control for improved ramping.[248]

*Demonstration Projects*

The experimental coupling of SOEC with methanol synthesis remains to be demonstrated and further research is required for large-scale application.

### *10.1.5 Discussion*

While extensive research has focused on the individual investigation of SOECs and downstream synthesis reactors, few studies have investigated the integrated behavior of entire P2X process chains, especially those including co-electrolysis. More work is needed on long-term performance, design concepts with optimized heat integration and dynamic operation.

Most downstream synthesis reactors can theoretically provide sufficient heat for SOEC water evaporation (Figure 14), as confirmed in different heat integration studies. However, large-scale systems must also consider additional heat demands, such as for $CO_2$ supply via CCS or DAC which so far have largely been excluded from analysis.

Regarding hydrogen/syngas purity, $H_2O$ is the primary inert gas and is typically unproblematic at lower concentrations as it already is a reaction byproduct in some of the discussed downstream processes. Where necessary, it can be removed via TSA/PSA, for instance, for ammonia synthesis. Nitrogen may enter the SOEC product gas stream in case of stack damage, leading to process flow adjustments, e.g. an increased purge gas ratio. Besides these inert gases, the influence of other potential impurities remains unclear, especially related to CO purity. For example, sulfur compounds present in $CO_2$ sources may initially adsorb on the SOEC fuel electrode, but could eventually breakthrough, leading to downstream catalyst degradation.

So far, steady-state SOEC-based P2X operation is preferred, however, fluctuating electricity prices from increasing RES penetration may necessitate more dynamic operation. Yet, downstream synthesis processes often have slower ramp rates and higher minimum loads (Table 4), making intermediate gas storage essential to align the dynamics of both sub-systems and to increase their capacity factors (Figure 16). Gas storage also allows to optimize the size of the synthesis units, which do not have to be designed for peak SOEC production. Size and cost of the gas storage system will be strongly dependent on the operational flexibility of the downstream reactors. In addition, the entire process chain needs to be designed to ensure the continuous availability of process heat for the SOEC steam generation, in particular during synthesis reactor downtime, which likely requires a heat management system. Large-scale plants often have on-site steam grids that can serve as thermal storage, while electric heaters can provide backup steam when process heat is insufficient.

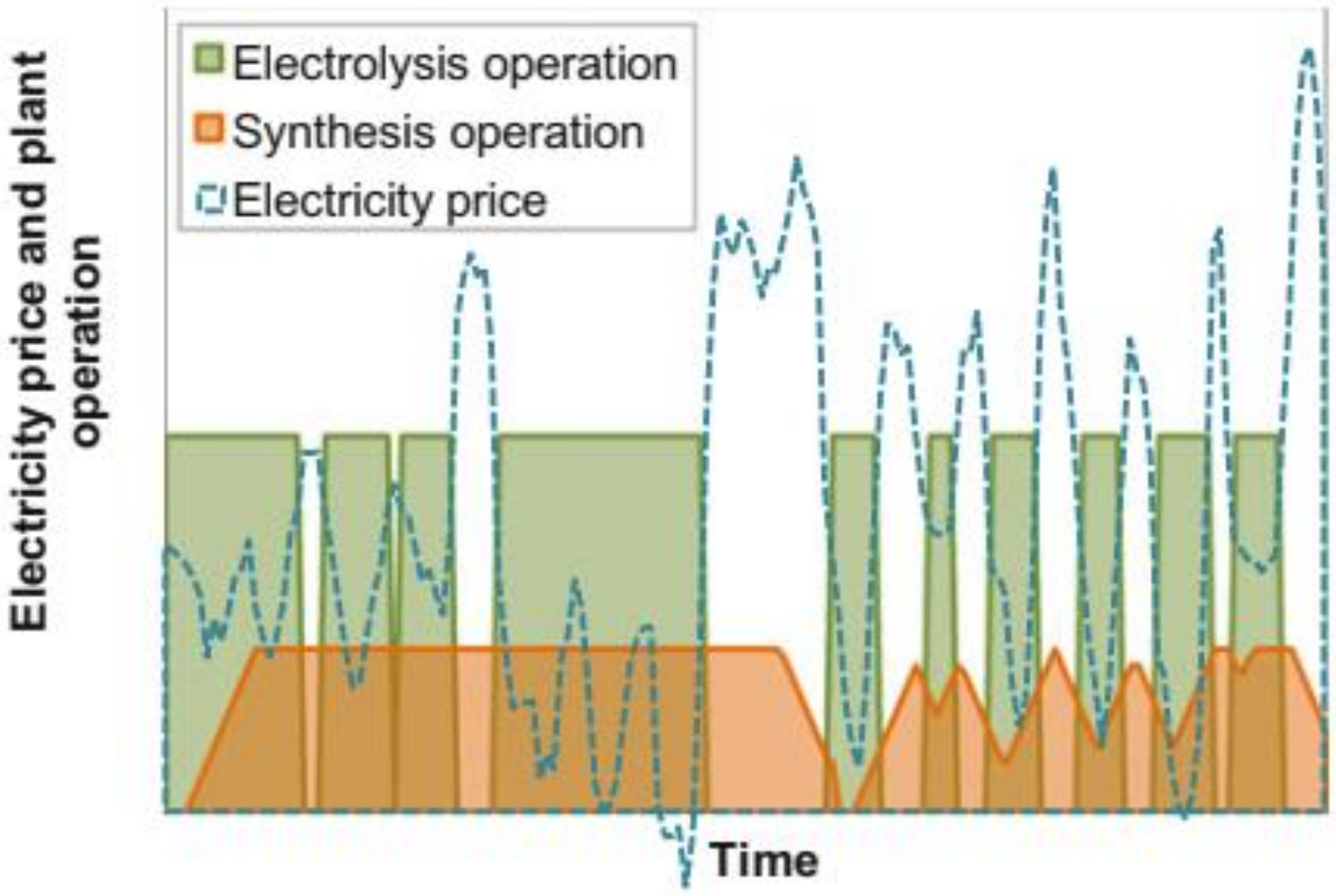

Figure 16. Operation strategy of a P2X process chain depending on an assumed electricity price. Reprinted with permission from Ref.[307]

Results of the first integrated systems with SOEC and downstream chemical processes will be key to validate P2X models and to assist scaling of systems up to the GW scale with optimized heat integration and buffer capabilities. Transient operation will require complex process control systems including the exchange of information between different plant modules such as storage level in the buffers, mass and energy flows. Based on high-detail models of the different plant components and forecasting of RES availability, digital twins with a continuous flow of data from the physical plant modules can optimize productivity for different time scales and capture the process chain dynamics.[308]

Large-scale SOEC coupling enables cost-effective, high-volume production and brownfield integration, while small, decentralized, flexible P2X units deserve further exploration for grid support and off-grid chemical production, especially where no distribution network is available.

### *10.2 Green steel production*

The iron and steel sector currently produces 7-9 % of global $CO_2$ emissions since the commonly used processes employ coal or coke for iron ore reduction via blast furnace routes.[241] While hydrogen, syngas, or methanol can be injected into blast furnaces with carbon capture, the emission reduction potential is limited to around 20%.[309] The primary renewable-based strategy for decarbonization of steel production is shifting most steelmaking to the direct reduced iron/electric arc furnace (DRI-EAF) route.[310] Direct reduction of iron ore is an established process that currently utilizes reformed natural gas as reducing agent in a shaft furnace to produce hot briquetted iron (HBI), already offering substantially reduced emissions compared to the blast furnace routes.[311]

The DRI stage can operate on 100% hydrogen or a hydrogen/natural gas mix,[312] which could serve as a bridging technology. Switching to green hydrogen for both high-temperature heat and reducing agents in the DRI-EAF process could reduce $CO_2$ emissions by up to 98% compared to blast furnace route.[241, 311] Direct reduction of iron ore with hydrogen is endothermic, whereas reduction via carbon monoxide is slightly exothermic:[309]

$$Fe_2O_3\,(s) + 3H_2(g) \rightarrow 2Fe(s) + 3H_2O(g) \qquad \Delta H\,(25°C) = 105\ kJ{\cdot}mol^{-1} \qquad (11)$$

$$Fe_2O_3\,(s) + 3CO(g) \rightarrow 2Fe(s) + 3CO_2(g) \qquad \Delta H\,(25°C) = -18\ kJ{\cdot}mol^{-1} \qquad (12)$$

The DRI stage is operated at ~800°C, with iron ore pellets pre-heated electrically before entering the furnace.[309] An over-stoichiometric hydrogen feed of $\lambda_{H2}$=1.5 is used,[309] and the operation pressure is 6-8 bar, with hydrogen compression contributing to its preheating. Metallization, defined as the percentage of total iron that exists in its metallic form, typically ranges from 90% to 95% in DRI.

Subsequently, the reduced iron leaves the shaft furnace and is briquetted in hot condition. As a next step, the HBI, steel scrap of up to 20 %, carbon and oxygen are introduced into the EAF to produce crude steel at ~1650°C.[313] Since the HBI still contains iron oxide and impurities which are reduced by external coal supply and a slag builder, an EAF off-gas is produced mainly containing of CO and $CO_2$.

Alternatively, the coupling of co-electrolysis with DRI-EAF can enhance the process by using carbon monoxide as a reducing agent and facilitating carburization in the DRI stage, which lowers the carbon demand in the EAF, improving energy efficiency and sustainability.[311, 314]

If pure oxygen is produced by the SOEC, it can be routed to a purge gas-oxygen burner for preheating the DRI feed gas or directly injected into the EAF. In the EAF, oxygen is crucial for decarburization, controlling the carbon content in steel, and in removing impurities such as silica, phosphorous. Additionally, it promotes foamy slag formation, which improves thermal efficiency by retaining heat and protecting the furnace lining during high-temperature operations.[309]

About one third of the total energy input of the EAF is carried away with the off-gas at high temperatures of ~1200°C, offering potential for heat integration. The heat could be used for scrap preheating,[315] or SOEC steam generation (Figure 17).[212, 311, 312] However, a heat storage, for instance a thermocline storage tank, would likely be necessary to smoothen out fluctuations since the EAF is a batch process.[315] Furthermore, it has been estimated that the EAF exhaust gas can supply only up to 12 % of the energy required for SOEC steam generation.[313] In contrast, the sensible heat contained in the liquid steel and slag is technically difficult to recover and therefore usually remains unused.

In DRI-EAF plants with integrated low-temperature electrolyzers, the DRI off-gas (~400°C) is cooled, water is condensed, a purge gas stream is separated and the remaining hydrogen is recirculated and heated back up in a heat exchanger. Subsequently, the recirculated gas is mixed with the fresh hydrogen from the electrolyzer and further heated up to the DRI operating temperature in two stages, using purge gas combustion and an electric heater. For SOEC, the heat exchanger after the DRI could be used for steam generation instead, with the off-gas reported to provide 47 % of the energy needed for water evaporation.[313] However, this increases the energy demand of the electric heater at the DRI inlet.

Overall, the integration of SOEC with DRI-EAF currently offers fewer synergies compared to the coupling with chemical synthesis processes, since most available heat streams are already utilized in the steelmaking process. However, DRI-EAF operation occurs at ambient pressure, which limits the potential benefits of using low-temperature electrolyzers that can operate at elevated pressures of up to 30 bar. As a result, SOEC remains the most efficient green hydrogen production route for steelmaking even with electrical water evaporation (Table 2), but unlocking its full efficiency potential depends on enabling thermal water evaporation. This represents a high-impact opportunity for process intensification that is still under exploration.

This pathway has gained considerable attention in steelmaking, for instance, the full integration of a 720 kW SOEC by Sunfire into the DRI-EAF operation site from Salzgitter AG (Germany) which aims to substitute 40 % of the natural gas supply with hydrogen.[228] Nevertheless, the fundamental process aspects of DRI-EAF plants operated on hydrogen, such as strategies for efficient heat integration and flexibility, remain insufficiently explored.

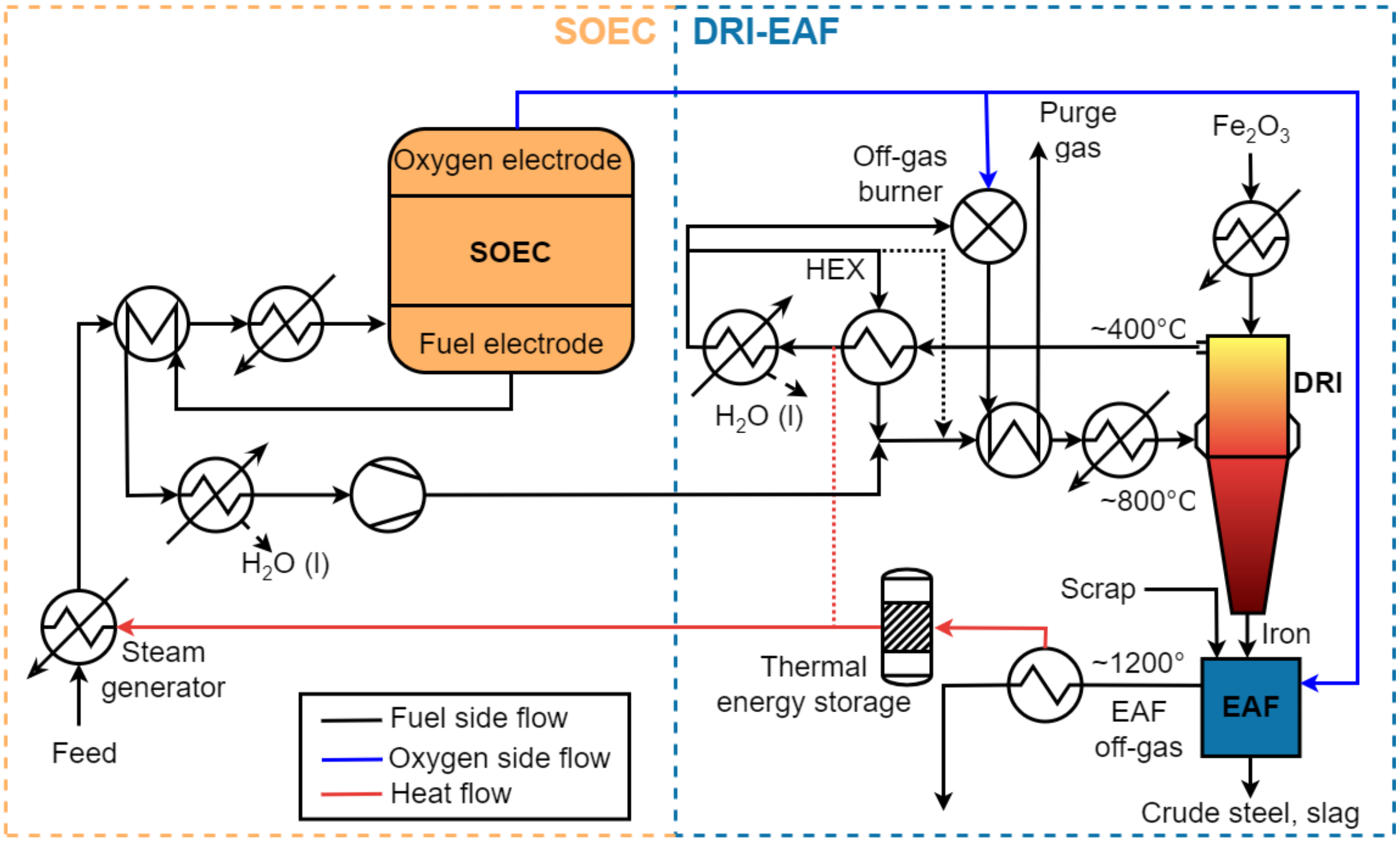

Figure 17. Simplified PFD for a DRI-EAF plant coupled to an SOEC showing the possible concept for heat and mass flow integration. The use of the excess heat from the EAF off-gas (red solid line) and from the DRI off-gas (dotted line) are depicted.

## 11 Summary and future R&D

SOEC technology has seen considerable advances in materials, cell and stack design in recent years and scale-up is now occurring rapidly. While continuous improvements in performance and degradation remain crucial, critical challenges for unlocking the full industrial potential of SOEC technology increasingly reside on the system level. This first system-level review of SOEC highlights the need for a paradigm shift from a materials-centric focus to a holistic, system-level approach. It identifies key barriers and proposes pathways to accelerate the deployment of SOEC systems as a cornerstone of the energy transition.

A key obstacle is the development of cost-effective and reliable BoP components such as electric heaters, heat exchangers and recirculation equipment. Pressurized systems that eliminate the need for purge air show promise for reducing CAPEX and enhancing overall efficiency. Despite increased operational challenges, co-electrolysis could also lead to CAPEX reductions and efficiency gains. However, while steam electrolysis has achieved TRL 7-8, co-electrolysis remains at TRL 5-6 and still requires demonstration at the MW scale. Although stable long-term performance with degradation rates below 0.5 %/kh can now reliably be achieved in both operating modes, further efforts are needed to reduce degradation, and more large-scale demonstrations for >10,000 h are required to reduce uncertainties regarding system lifetime and de-risk investments.

SOEC systems are currently designed for isothermal steady-state operation for cost, efficiency and durability considerations. However, the increasing share of RES in the energy system will increase the necessity for flexible and dynamic operation. This review has highlighted that SOECs possess better dynamic operating capabilities than the most prominently discussed exothermic downstream synthesis processes. Moreover, advancements in control and operating strategies such as MPC, modular operating concepts and PWM can further mitigate degradation and ensure long-term performance. To address high CAPEX, the development of larger cells, stacks and stack modules is required, with the SOEC areal footprint already decreasing rapidly. Effective modularization concepts will be vital for a cost-effective and reliable operation of such systems. As SOEC technology is up-scaled, the design and demonstration of systems with an increased level of integration with upstream and downstream processes increases in importance. Early demonstration systems comprising entire P2X process chains including $CO_2$ capture, SOEC and downstream chemical processes will provide essential insights to design GW scale systems with optimized heat integration and buffer capabilities.

This review emphasized the interdisciplinary challenges and transformative potential of SOEC systems, positioning them as a critical enabler for decarbonizing industrial processes and facilitating the global energy transition.

**Acknowledgements**

The authors are grateful to Andreas Mai from Topsoe A/S for discussions about industrial scale-up challenges and scientific input.